\documentclass[12pt, twocolumn, amsmath,amssymb]{iopart}
\usepackage{iopams}
\usepackage{widetext}
\usepackage{rotating}
\usepackage{afterpage}
\usepackage{color}
\usepackage{graphicx}% Include figure files
\usepackage{dcolumn}% Align table columns on decimal point
\usepackage{bm}% bold math

\begin{document}

\title[Effective Field Theory of Chiral Spin Liquid between Ordered Phases in Kagom\'{e} Antiferromagnet]{Effective Field Theory of Chiral Spin Liquid between Ordered Phases in Kagom\'{e} Antiferromagnet}

\author{Imam Makhfudz}

\address{Univ Lyon, Ens de Lyon, Univ Claude Bernard, CNRS, Laboratoire de Physique, F-69342 Lyon, France}
\ead{imakhfudz@gmail.com}
\vspace{10pt}
\begin{indented}
\item[]August 2017
\end{indented}

\begin{abstract}
We propose in this work an effective field theory description of the chiral spin liquid state in Heisenberg spin system on kagom\'{e} lattice.To this end, we derive the  low-energy effective theory of kagom\'{e} (isotropic) Heisenberg antiferromagnet around its ordered ground states found numerically and show that quantum fluctuations induced by further neighbor spin exchanges are equally strong as those from first neighbor.We use a chiral order parameter theory to argue for the occurrence of finite temperature chiral symmetry breaking transition into chiral ordered state in kagom\'{e} antiferromagnet with further neighbor spin exchange interactions.We compute the chiral symmetry breaking term in the effective ground state energy and show that chiral spin liquid necessarily occurs in the ground state of kagom\'{e} antiferromagnet with the first three nearest-neighbor spin exchange interactions.Finally, we consider the quantum criticality of the kagom\'{e} antiferromagnet and show that a Chern-Simons term emerges naturally at the transition between two ordered states that satisfy appropriate `matching condition' that we derive, providing explanation for why chiral spin liquid could occur at the transition between appropriate ordered states.This emergent Chern-Simons term is the low energy effective theory of the chiral spin liquid state, where the chirality is the immediate consequence of the breaking of discrete symmetries by this topological field theory.
\end{abstract}

%\maketitle

\section{Introduction}Quantum spin liquid has been a highly sought-after intriguing quantum state of matter for decades, ever since its first conception as a possible new type of insulating ground state by Phil Anderson in 1973 in his study of antiferromagnet in triangular lattice \cite{PhilAnderson1973}.The discovery of high $T_c$ superconductivity in cuprates prompted another impetus for interest in this state of matter due to the proposal that this state might arise in the physics of this strongly correlated superconducting material \cite{LeeNagaosaWen}.The search for spin liquid expanded to magnetic materials especially on frustrated spin systems where geometry and spin exchange interaction collaborate to give rise to ground state degeneracy and quantum fluctuations that tend to disorder the ground state \cite{Balents}.The search for spin liquid state is still very intensive to these days in both theoretical and experimental fronts \cite{SavaryBalents}.

Frustrated spin systems have provided promising candidates to host quantum spin liquid state.We will focus on one of such frustrated spin systems in this work, namely kagom\'{e} antiferromagnet, where the spins live on lattice with corner-sharing triangles as shown in Fig.1.Candidate materials with kagom\'{e} lattice structure have been discovered in the past few years that open the opportunity to explore the existence of quantum spin liquid state in those materials \cite{MNorman}.The presence of triangles and antiferromagnetic spin exchange interaction make this system highly frustrated.The ground state of kagom\'{e} has been the subject of intense studies for several decades.Some works proposed that the ground states are ordered \cite{Huse}\cite{Chalker}\cite{Sachdev1992} while some others, especially those from recent years, proposed that the strong frustration may actually permit quantum disordered ground state; the quantum spin liquid.The precise of nature of such quantum spin liquid state has also been an intense debate with significant numerical works in the past a decade suggested spin liquid state that takes the form of U(1) Dirac spin liquid \cite{RanHermeleLeeWen}\cite{Yasir}, gapped $Z_2$ spin liquid with topological order \cite{White}\cite{Schollwock}, and chiral spin liquid \cite{Messio}.We will be concerned especially on chiral spin liquid in this work, motivated by the observation that such state would manifest a spin analog of fractional quantum Hall effect \cite{Tsui}, the theoretical explanation of which \cite{Laughlin} has contributed to the birth of the research on topological phases of matter.Chiral spin liquid is a type of quantum spin liquid that breaks some type of discrete symmetry associated with chirality, as defined by the scalar spin chirality 
\begin{equation}\label{SSC}
\chi_{ijk}=\mathbf{S}_i\cdot(\mathbf{S}_j\times\mathbf{S}_k)
\end{equation}
for any three spins on a triangular plaquette of the lattice in which spins live \cite{WenWilczekZee}.Chiral spin liquid can be characterized by long range order of such chirality but in the absence of long range spin order.Other characterization involves Laughlin type of wave function motivated by the analogy between chiral spin liquid physics and the physics of fractional quantum Hall effect \cite{KalmeyerLaughlin}\cite{KunYangGirvin}.There have been extensive theoretical and numerical studies on this form of spin liquid over decades on simple model as well as truly frustrated spin systems up to recently, especially on the popular candidate system believed to host quantum spin liquid state; the kagom\'{e} antiferromagnet, with isotropic Heisenberg antiferromagnetic spin exchange interaction \cite{DNShengSciRepFQHEkagome}, anisotropic version \cite{DNShengPRLchiralSL}\cite{YCHe}, as well as the one that also includes explicit scalar spin chirality term \cite{Bauer}\cite{Lauchli}, but a full conceptual understanding is yet to be achieved and the very existence and characterization of the properties of this chiral spin state are to be clarified both theoretically and experimentally.

We present in this work a low-energy effective theoretical point of view on chiral spin liquid ground state implemented directly to kagom\'{e} antiferromagnet with isotropic Heisenberg spin exchange interactions.The crucial insight provided by an effective theory is that chiral spin liquid arises from chiral symmetry breaking effect described by an action involving Chern-Simons term, which breaks discrete symmetries in the form of time-reversal and parity symmetries \cite{WenWilczekZee}\cite{FradkinSchaposnik}.An \textit{induced} Chern-Simons term in 2d antiferromagnet has been proposed recently \cite{ImamPierre} in a slightly different context of doped spin system, but an explicit demonstration of \textit{spontaneous} emergence of such topological term beyond symmetry argument in the context of chiral spin liquid in pure spin systems is a challenging task.An attempt to formulate a low energy effective theory of chiral spin liquid has been made for an anisotropic version of kagom\'{e} antiferromagnet and one that includes explicit scalar spin chirality within Schwinger boson quantum formalism \cite{YCHePollmannMoessner}.We provide in this work a simpler and yet transparent formulation of field theory of chiral spin liquid on isotropic kagom\'{e} Heisenberg antiferromagnet with no explicit scalar spin chirality and show how Chern-Simons term can naturally emerge in this system spontaneously.   

We consider isotropic Heisenberg Hamiltonian defined on kagom\'{e} lattice with first, second, and third neighbor spin exchange interactions,
\[
H=\sum_{ij\in nn,nnn,nnnn}J_{ij}\mathbf{S}_i\cdot\mathbf{S}_j
\]
\begin{equation}\label{HeisenbergAFMkagome}
=\sum_{ij\in nn}J_1\mathbf{S}_i\cdot\mathbf{S}_j+\sum_{ij\in nnn}J_2\mathbf{S}_i\cdot\mathbf{S}_j+\sum_{ij\in nnnn}J_3\mathbf{S}_i\cdot\mathbf{S}_j
\end{equation}
with first ($nn$), second ($nnn$), and third ($nnnn$) neighbor spin exchange interactions with couplings $J_{nn}=J_1>0,J_{nnn}=J_2,J_{nnnn}=J_3$ respectively.In the rest of the paper, we use the phrase `kagom\'{e} antiferromagnet' to represent our choice that the first neighbor spin exchange coupling is antiferromagnetic $J_1>0$.In \cite{Messio}, numerical calculation based on Schwinger boson mapping of $S=1/2$ Heisenberg model with $J_1=1$ and ferromagnetic ($J_2,J_3<0$) or antiferromagnetic ($J_{2,3}>0$) further neighbor spin exchanges suggested the occurrence of gapped chiral topological spin liquid, which put forward alternative scenario deduced from earlier numerical works which found gapped $\mathbb{Z}_2$ spin liquid in this system \cite{White}\cite{Schollwock}.This very motivating result of \cite{Messio} was further corroborated by other numerical studies also found chiral spin liquid in kagom\'{e} antiferomagnet \cite{DNShengSciRepFQHEkagome}\cite{DNShengPRLchiralSL}\cite{Globalphasediagram}\cite{Wenjun}.The wavefunction overlap of the chiral spin liquid ground state with projected Gutzwiller wave function was analyzed and nice overlap was found \cite{Lauchli}.Notably, it was argued in \cite{Lauchli} in conjunction with \cite{Messio} and \cite{DNShengSciRepFQHEkagome} that the chiral spin liquid occurs only between the nonplanar ordered state with 12 sublattices called `cuboc1' \cite{Paris} and the planar ordered state called $\mathbf{q}=0$ first proposed in \cite{Sachdev1992}.One of the goals of this work is to provide a universal field theoretical explanation for this numerical conclusion.We accomplish this by considering low-energy effective theory of the spin system around the ordered phases that dominate the ground state of kagom\'{e} Heisenberg model.

We will first consider the theoretical possibility of quantum spin liquid state in kagom\'{e} antiferromagnet described by Eq.(\ref{HeisenbergAFMkagome}).In order to have a spin liquid state, one needs strong quantum fluctuations that could destroy the magnetic ordered ground states at $T=0$ associated with the frustration effect which is present naturally in kagom\'{e} antiferromagnet due to antiferromagnetic spin exchange interaction between the spins on the triangles that make up the kagom\'{e} lattice.The consideration of only first neighbor spin exchange in existing literature (e.g. \cite{White}) is reasonable since it is the strongest and dominant spin exchange interaction in kagom\'{e} antiferromagnet.Being the strongest spin exchange interaction, one would naively expect that the first neighbor spin exchange interaction would also give dominant quantum fluctuations effect that is needed to give rise to the quantum spin liquid state.Further neighbor spin exchange interactions are naively expected to be much weaker, both in its strength and the resulting quantum fluctuations, and so are expected to give no qualitative change in the physics compared to that from the nearest neighbor spin exchange interaction.

We will show in this work that the above naive expectations are not in fact the prevailing case.First, we will show that, despite the further distances and supposedly weaker strength, the geometric and configurational pattern of the spin ordered states can render the further neighbor spin exchange interactions be able to give rise to equally strong quantum fluctuations as those due to the first (nearest neighbor) spin exchange interaction.In particular, in an explicit calculation on $J_1-J_2-J_3$ kagom\'{e} antiferromagnet model, we found that the quantum fluctuations due to the third neighbor spin exchange are of the same order of magnitudes as those due to the first neighbor, while being much greater that those due to the second neighbor spin exchange interaction.Second, we will show that while kagom\'{e} antiferromagnet with only first neighbor spin exchange interaction has an effective local $\mathbb{Z}_2$ symmetry that prohibits finite temperature phase transition, further neighbor spin exchange interactions reduce this effective local $\mathbb{Z}_2$ symmetry to that of global Ising symmetry that allows finite temperature phase transition.This transition is chiral phase transition that gives rise to chiral spin state at finite temperature, below the critical temperature of the phase transition.This chiral spin state opens up the possibility for chiral spin liquid ground state.The key to such chiral spin liquid state is the emergence of Chern-Simons term, which is an effective field theoretical description of chiral spin liquid state.We show explicitly that this Chern-Simons term appears naturally at the phase boundary between appropriate ordered states satisfying a criterion, which is the third key result of the paper.

The implications of the above results are as follows.The first result suggests that it is necessary to consider further neighbor spin exchange interactions in kagom\'{e} antiferromagnet as they give rise to equally strong quantum fluctuations and change the nature of ground state dramatically.In particular, the first two results explain why numerical results found chiral spin liquid for kagom\'{e} antiferromagnet with further neighbor spin exchange interaction (e.g. \cite{Messio}\cite{Globalphasediagram}\cite{Wenjun}) while those with only first neighbor found time-reversal and parity symmetric (e.g. \cite{White}) types of spin liquid states.Our third result shows that chiral spin liquid state occurs at the quantum critical phase boundary between two appropriate ordered states satisfying appropriate matching criterion, which we will show to agree fully with the existing numerical results in literature. 

\begin{figure}
\begin{center}
 \includegraphics[angle=0,origin=c, scale=0.10]{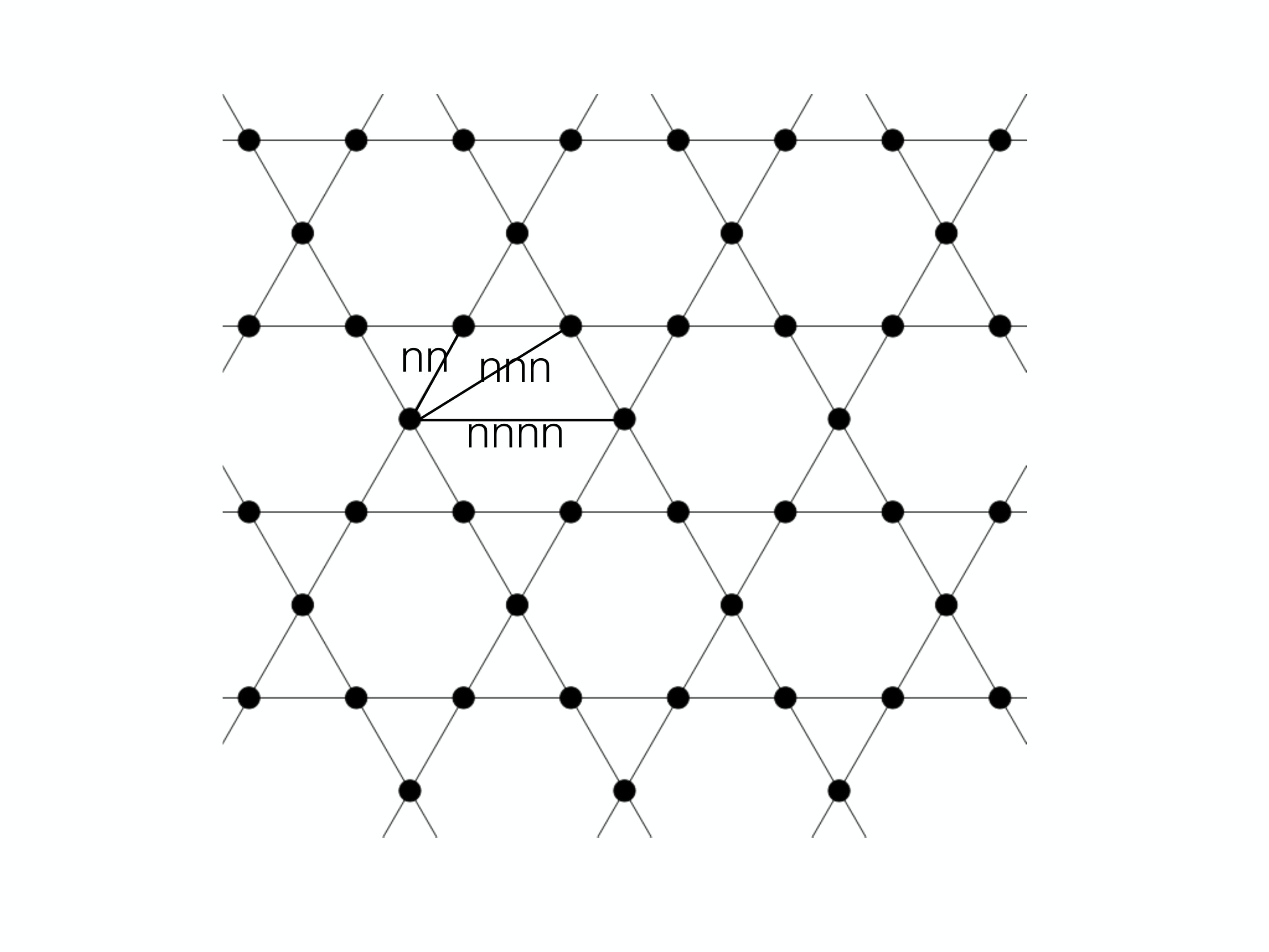}
 \label{fig:kagomeAFM}
 \caption{
 The lattice structure of kagom\'{e} antiferromagnet with the first (nn), second (nnn), and third (nnnn) neighbor spin exchange interactions.}
 \end{center}
 \end{figure}

\section{The Ordered Ground States and Their Low Energy Effective Theories}

Our original goal is to obtain an effective theory that describes a quantum disordered spin state that preserves spin rotational symmetry.Ideally, a fully quantum description that treats spin as a quantum mechanical operator right from the beginning is desired.Following existing literature, one can adopt an appropriate slave-particle formalism that expresses a spin operator in terms of quantum particle operator coupled to an appropriate gauge field.A mean-field theory treatment of the resulting quantum Hamiltonian is expected to describe magnetically ordered or quantum disordered spin state.An effective field theory that describes the latter case can be obtained by expanding the corresponding action in terms of the gauge field degree of freedom around the mean-field state.Rather than pursuing this nontrivial task, we will instead take advantage, as a starting point in this theoretical work, of the numerical observations that the ground state of kagom\'{e} antiferromagnet is dominated by magnetically ordered states \cite{Messio}\cite{Domenge}.Since we have ordered states, we can use semiclassical treatment of spin systems where the spin is treated as a classical spin vector.

A general field theory that can describe all possible types of ground state, be they ordered or disordered, can be derived by evaluating the exact partition function of the system from which one obtains the effective action in general case.The action (or free energy) is then minimized at `mean-field level' and one obtains a particular ordered or disordered ground state depending on the microscopic parameter.The corresponding effective action describing the quantum fluctuations can be obtained by rewriting the order parameter field as a sum of mean-field part and fluctuating part describing the quantum fluctuations, normally represented by a gauge field.The effective action for the fluctuating field is obtained by summing over all possible spin configurations at the mean-field level, that is, performing a path integral, and expressing the partition function in terms of the remaining gauge degree of freedom of interest.This is again an analytically challenging task but what we will do here is to derive a low-energy effective Hamiltonian by directly expanding the original microscopic spin Hamiltonian in terms of the gauge degree of freedom around those ordered states \cite{Haldane}.The resulting effective Hamiltonian is expected to give the same physics as the path integral-derived effective action because, in addition that Hamiltonian and Lagrangian are directly related by Legendre transformation, both approaches are based on variational principle; minimization of the action in the path-integral approach and minimization of energy in the Hamiltonian approach with the latter assumes a particular ground state that is obtained from the minimization of the action in the former approach.

Such low energy effective theory describes the dynamics of the fluctuations around the ordered state spin configuration corresponding to Goldstone modes.Normally, these fluctuations are well understood to be the thermal fluctuations induced at finite but low temperatures around the zero temperature ground states.We assert that the low energy effective theory that we will derive below can also describe the quantum fluctuations at zero temperature that give rise to a quantum spin disordered ground state.We put forward the following arguments for the justification of this assertion.Firstly, the supposed quantum spin disordered state is expected to occur in the vicinity of the ordered states, that is, to be precise, between ordered ground states.This situation has indeed been observed numerically \cite{Lauchli}.Therefore, since the appropriate low energy effective theories can be derived for the ordered ground states on the two sides, one should expect that an appropriate low energy effective theory should be derivable as an interpolation between the two effective theories for the ordered states, with expected nontrivial terms that give rise to disorderedness.Secondly, a quantum phase transition \cite{Sachdev} can be perceived as a classical (thermodynamic) phase transition where the temperature axis is replaced with a coupling parameter at zero temperature.Therefore, the low energy effective theory that describes a classical phase transition from an ordered ground state to a classical (finite temperature) paramagnetic state should perceivably be connectible to a field theory that describes a quantum phase transition from an ordered ground state to a quantum paramagnetic state at zero temperature.In other words, the temperature $T$ in the standard expression for partition function is replaced by an appropriate coupling parameter that drives the kinetics at zero temperature \cite{Makhfudz}.Thirdly and perhaps the most fundamental one, the low-energy effective theory describing the quantum fluctuations that give rise to disordered ground state preserves the global symmetry of the original microscopic Hamiltonian as we will see later where the spin rotational symmetry of the original Hamiltonian manifests as a gauge invariance in the low-energy gauge field theory.That an effective field theory preserves the symmetry of the original microscopic Hamiltonian is of course a mandatory condition for the former to be valid.

The above physical picture is intuitively very sensical and consistent with basic principles to say the least.With that picture in mind, one can then define an appropriate gauge field that defines the effective gauge degree of freedom to represent the quantum fluctuations around the ordered ground state, which is thus one step forward beyond mean-field description.This gives rise to an emergent gauge theory which can be analyzed via elegant duality transformation in terms of dual theory \cite{TanakaTotsukaHu}.The results to be derived in this and following sections seem to support the validity of our semiclassical field theory description of the quantum spin disordered state obtained from the expansion around the neighboring ordered ground states, made possible by the physical intuitiveness of the above assertion.

Numerical works based on Schwinger boson mean field theory \cite{Messio} found a ground state phase diagram with cuboc1, $\mathbf{q}=0$, and $\sqrt{3}\times\sqrt{3}$ ordered states.We will analyze these ordered ground states within our low-energy effective theory picture.The $\sqrt{3}\times\sqrt{3}$ and $\mathbf{q}=0$ states are coplanar ordered states with $120^{\mathrm{o}}$ spin structure where the three spins on each triangle of the kagom\'{e} lattice form a mercedes star and effective flux from curling chirality with respectively alternating sense (clockwise vs. anticlockwise) and non-alternating clockwise sense on each two adjacent triangles around the David star \cite{Sachdev1992}.The cuboc1 state on the other hand is a non-coplanar ordered state with out-of-plane spin vectors within 12 sublattices per unit cell, with the twelves spins point toward the 12 corners of a cubochedron.Within semiclassical approach, the spin is represented as classical vector $\mathbf{S}_i=S(\sin\theta_i\cos\phi_i,\sin\theta_i\sin\phi_i,\cos\theta_i)$ from which the Heisenberg Hamiltonian Eq.(\ref{HeisenbergAFMkagome}) becomes
\begin{widetext}
\[
H=\sum_{ij}S^2J_{ij}\left(\cos\theta^0_i\cos\theta^0_j+\cos\Delta\phi^0_{ij}\sin\theta^0_i\sin\theta^0_j\right)
\]
\[
\fl +
S^2J_{ij}\left[-\sin\Delta\phi^0_{ij}\sin\theta^0_i\sin\theta^0_j\delta\phi+(\sin\theta^0_i\cos\theta^0_j\cos\Delta\phi^0_{ij}-\cos\theta^0_i\sin\theta^0_j)\delta\theta\right]
\]
\begin{equation}\label{Hamiltonianfluctuation}
\fl +S^2J_{ij}\left[-\frac{1}{2}\cos\Delta\phi^0_{ij}\sin\theta^0_i\sin\theta^0_j(\delta\phi^2+\delta\theta^2)-\sin\theta^0_i\cos\theta^0_j\sin\Delta\phi^0_{ij}\delta\phi\delta\theta-\frac{1}{2}\cos\theta^0_i\cos\theta^0_j\delta\theta^2\right]
\end{equation}
\end{widetext}
where $\Delta\phi^0_{ij}=\phi^0_j-\phi^0_i$ and working to first order in the gradients of the angular fields, we have $\delta\theta\equiv\delta\theta_{ij}=\Delta\mathbf{r}_{ij}\cdot\nabla\theta+\Delta \tau \partial_{\tau}\theta+\mathcal{O}((\nabla\theta)^2,(\partial_{\tau}\theta)^2,\nabla\theta\partial_{\tau}\theta)$ and $\delta\phi\equiv\delta\phi_{ij}=\Delta\mathbf{r}_{ij}\cdot\nabla\phi+\Delta \tau \partial_{\tau}\phi+\mathcal{O}((\nabla\phi)^2,(\partial_{\tau}\phi)^2,\nabla\phi\partial_{\tau}\phi)$ where $\tau$ is imaginary time as we are working in Euclidean space-time.The next step is to determine an appropriate definition of gauge field in terms of $\nabla\theta$ and $\nabla\phi$.We define the gauge field as 

\begin{equation}\label{gaugefielddefinition}
A_{\mu}(\mathbf{r})=\mathbf{A}^{\mathrm{monop}}(\mathbf{r})\cdot\partial_{\mu}\mathbf{n}(\mathbf{r})
\end{equation}
where $\mu=0,1,2$, $\mathbf{n}$ is local unit vector giving the direction of magnetization $\mathbf{n}_i=\mathbf{S}_i/S$ and $\mathbf{A}^{\mathrm{monop}}$ is monopole vector potential in appropriate gauge \cite{Shankar}.We find that, the resulting vector potential is
\begin{equation}\label{gaugefield}
\mathbf{A}(\mathbf{r})=\left(1-\cos\theta(\mathbf{r})\right)\nabla\phi(\mathbf{r})
\end{equation}
in the continuum limit and corresponds to a gauge choice where the monopole vector potential is written as $\mathbf{A}^{\mathrm{monop}}(\mathbf{r})=(1-\cos\theta(\mathbf{r}))/\sin\theta(\mathbf{r})\hat{e}_{\phi}$ while gauge transformation simply adds the above $\mathbf{A}(\mathbf{r})$ by $\delta\mathbf{A}(\mathbf{r})=n\nabla\phi(\mathbf{r})$, where $n$ is arbitrary integer, which does change the effective magnetic field $\mathbf{B}$, according to the identity $\oint_C d\mathbf{r}\cdot \mathbf{A}=\int d\mathbf{S}\cdot \mathbf{B}$, where $\mathbf{B}=\nabla\times\mathbf{A}$ and the line integral is along the closed loop $C$ that encloses the surface $\mathbf{S}$.The invariance under the gauge transformation $\mathbf{A}\rightarrow \mathbf{A}+n\nabla\phi$ can be easily seen by considering a closed loop enclosing any set of lattice sites and as one integrates $\int d\mathbf{r}\cdot n\nabla\phi$ along the loop, the result is $2\pi n$, which is equivalent to zero effective magnetic flux.We have also checked that using CP$^1$ representation \cite{Fradkinbook} gives the same result for planar spin ordered state $A_{\mu}=\partial_{\mu}\phi$ and our definition based on Eq.(\ref{gaugefielddefinition}) works even for nonplanar spin ordered state.

For planar ordered states such as the $\mathbf{q}=0$ and $\sqrt{3}\times\sqrt{3}$ states, $\cos\theta(\mathbf{r})=0$ and we thus have simple expression $\mathbf{A}(\mathbf{r})=\nabla\phi(\mathbf{r})$, in agreement with that in \cite{ImamPierre}.It is to be noted also that to first order in ($\nabla\phi,\nabla\theta$) the emergent gauge field is dependent only on the spatial gradient of the azimuthal angle $\phi$ because, interestingly, the $\nabla\theta$-dependent terms cancel out exactly at this order.This is remarkable observation and suggests that since we are considering 2d system, even though the spin vector lives in SO(3) space, only the 2d projection of the spin vector dynamics would define the effective gauge field, which should indeed live in the 2d space on which the spin system lives.In connection to this, the $\delta\theta$-dependent terms in the low-energy Hamiltonian Eq.(\ref{Hamiltonianfluctuation}) define a kinetic momentum operator corresponding to a much more massive mode (due to the very last term in Eq.(\ref{Hamiltonianfluctuation})) which ought to be integrated out when we are interested in the low energy effective gauge field Hamiltonian \cite{ImamPierre} to be derived in this section.Formally, this is done using functional integral formalism by considering the partition function in terms of the fluctuating fields $\delta\phi,\delta\theta$ given by
\begin{equation}\label{derivation}
\fl Z=\int \mathcal{D}\delta\phi\int \mathcal{D}\delta\theta e^{-S[\delta\phi,\delta\theta]}=\int \mathcal{D}\delta\phi\int \mathcal{D}\delta\theta e^{-\int d\tau[\zeta^T(\partial_{\tau}+M)\zeta]+K_{\phi}\delta\phi+K_{\theta}\delta\theta}
\end{equation} 
where we work in Euclidean space-time (with imaginary time $\tau$), $\zeta=(\delta\phi,\delta\theta)^T$, $S[\delta\phi,\delta\theta]=\int d\tau [\zeta^T\partial_{\tau}\zeta+H(\delta\phi,\delta\theta)]$ and the $2\times 2$ matrix $M$ and the constants $K_{\phi},K_{\theta}$ can be read off from Eq.(\ref{Hamiltonianfluctuation}).Doing the Gaussian integration out of the $\delta\theta$, we arrive at the following effective Hamiltonian from Eqs.(\ref{Hamiltonianfluctuation}) and (\ref{derivation}) 
\begin{widetext}
\[
H=H_0-\frac{1}{2}S^2\sum_{ij}J_{ij}\cos\Delta\phi^0_{ij}\sin\theta^0_i\sin\theta^0_j\left[(\nabla\phi\cdot\Delta\mathbf{r}_{ij})^2+(\Delta\tau\partial_{\tau}\phi)^2\right]
\]
\begin{equation}\label{lowenergyHamiltonian}
= H_0-\frac{1}{2}S^2\sum_{ij}J_{ij}\cos\Delta\phi^0_{ij}\mathrm{cot}\frac{\theta^0_i}{2}\mathrm{cot}\frac{\theta^0_j}{2}\left[(\Delta\mathbf{r}_{ij})^2(\mathbf{A}^{\parallel})^2+(\Delta\tau)^2A^2_0\right]
\end{equation}
\end{widetext}
where we have used the cancellation of linear terms by symmetry, used the gauge field expression Eq.(\ref{gaugefield}) to first order in ($\nabla\phi,\nabla\theta$) by using the mean-field equilibrium value $\theta^0$ in place of $\theta(\mathbf{r})$ and defined 
\begin{equation}
\mathbf{A}^{\parallel}=(\mathbf{A}\cdot\Delta\mathbf{r}_{ij})\frac{\Delta\mathbf{r}_{ij}}{|\Delta\mathbf{r}_{ij}|^2}.
\end{equation}
with renormalized coupling constants $J_{ij}$ and from Eq.(\ref{Hamiltonianfluctuation}), we have
\begin{equation}\label{GSenergy}
H_0=\sum_{ij}S^2J_{ij}\left(\cos\theta^0_i\cos\theta^0_j+\cos(\phi^0_j-\phi^0_i)\sin\theta^0_i\sin\theta^0_j\right)
\end{equation}
which gives the ground state energy $E$ determining the energetics of the ordered state.This ground state energy has been computed in \cite{MessioPRBregularmagneticorder} and is consistent with what one will get from the above Eq.(\ref{GSenergy}).We will mainly be concerned with the quantum fluctuations part represented by the gauge field terms in Eq.(\ref{lowenergyHamiltonian}).The appearance of explicit mass term for the gauge field in Eq.(\ref{lowenergyHamiltonian}) represents a dynamical mass generation for the gauge field.That is, the Anderson-Higgs mechanism, which is a well-established concept in quantum field theory \cite{Pokorski}.Furthermore, in terms of the original spin model Eq.(\ref{HeisenbergAFMkagome}), a gauge transformation corresponds to a spin rotation.What this means is that the Hamiltonian Eq.(\ref{HeisenbergAFMkagome})) is invariant under global spin rotation, that is, the rotation of all spins in the lattice by the same amount about arbitrary axis, and also local rotation of any one of the spins by integer multiples of $2\pi$ around the $z$ axis, both of which of course have no physical consequences, just like a gauge transformation.According to Eq.(\ref{gaugefield}) and the following discussion above, the change in azimuthal angle $\phi$ corresponds to a gauge transformation $\mathbf{A}\rightarrow \mathbf{A}+n\nabla\phi$, which does not change the effective magnetic flux $\mathbf{B}=\nabla\times\mathbf{A}$.The starting model Eq.(\ref{HeisenbergAFMkagome}) has a global spin rotational symmetry corresponding to a global gauge invariance, which is still preserved by the gauge Hamiltonian Eq.(\ref{lowenergyHamiltonian}) because for global gauge transformation, $\delta A_{\mu}\sim\partial_{\mu}\phi=0$.Therefore, our gauge field theory Eq.(\ref{lowenergyHamiltonian}) respects the global symmetry of the starting spin model Eq.(\ref{HeisenbergAFMkagome}).The local gauge invariance corresponding to rotation of any one of the spins by $2\pi n$ around the $z$ axis possessed by the original Hamiltonian Eq.(\ref{HeisenbergAFMkagome}) is no longer present in the effective Hamiltonian and need not indeed be present because Eq.(\ref{lowenergyHamiltonian}) was derived around a symmetry breaking ordered state.  

For planar spin ordered states, we have $\theta^0_i=\pi/2$ for all $i$ so that
%\begin{widetext}
\begin{equation}\label{gaugefieldHamiltonianq=0}
H=H_0-\frac{1}{2}S^2\sum_{ij}J_{ij}\cos\Delta\phi^0_{ij}\left[(\Delta\mathbf{r}_{ij})^2{\mathbf{A}^{\parallel}}^2+(\Delta\tau)^2A^2_0\right]
\end{equation}
%\end{widetext}
where we have used from Eq.(\ref{gaugefield}) that $\mathbf{A}(\mathbf{r})=\nabla\phi(\mathbf{r})$ for planar ordered states.The derivation of the continuum limit of the effective Hamiltonian from Eq.(\ref{lowenergyHamiltonian}) involves a very tedious labor of calculating the geometric factors of the spin configurations considered in \cite{Messio} but otherwise quite straightforward.Using the observation that for $\sqrt{3}\times\sqrt{3}$ state $\Delta\phi^0_{ij\in nn,nnn,nnnn}=2\pi/3,0, 2\pi/3$ and for $\mathbf{q}=0$ state $\Delta\phi^0_{ij\in nn,nnn,nnnn}=2\pi/3,2\pi/3,0$, we obtain
%\begin{widetext}
\begin{equation}\label{gaugefieldHamiltonian3x3result}
H=H_0+\frac{1}{2}\frac{S^2}{\mathcal{A}}(J_1+6|J_2|+2J_3)\int d^2\mathbf{r}({\mathbf{A}^{\parallel}}^2(\mathbf{r})+A^2_0(\mathbf{r}))
\end{equation}
for the $\sqrt{3}\times\sqrt{3}$ state, where we have used the fact that this state only occurs for $J_2<0$ at the classical level \cite{Messio} and
\begin{equation}\label{gaugefieldHamiltonianq=0result}
H=H_0+\frac{1}{2}\frac{S^2}{\mathcal{A}}(J_1+3J_2-4J_3)\int d^2\mathbf{r}({\mathbf{A}^{\parallel}}^2(\mathbf{r})+A^2_0(\mathbf{r}))
\end{equation}
%\end{widetext}
for the $\mathbf{q}=0$ state, where $N=\int d^2\mathbf{r}/\mathcal{A}$ is the number of sites of kagom\'{e} lattice with unit cell area $\mathcal{A}$ used with the continuum limit $\sum_{i}\rightarrow \int d^2\mathbf{r}/\mathcal{A}$ and we have made use of the fact that given unit lattice spacing $a=1$, we then have $|\Delta\mathbf{r}_{nn}|=1,|\Delta\mathbf{r}_{nnn}|=\sqrt{3},|\Delta\mathbf{r}_{nnnn}|=2$ and we have chosen the $\Delta\tau$ appropriately given by the energy scale determined by the spatial part.An important observation of the result Eq.(\ref{gaugefieldHamiltonianq=0result}) for the $\mathbf{q}=0$, which is a coplanar spin ordered state, the relative weight of the first and further neighbor contributions is of the same order of magnitude.This is related to the fact that for coplanar state, the scalar spin chirality $\chi_{ijk}$, when the spins are treated as classical vectors, vanishes identically, so that there is no difference between the first and further neighbor spin exchanges.

For the cuboc1 state, we use $\mathbf{A}^2=(1-\cos\theta^0_i)\nabla\phi(1-\cos\theta^0_j)\nabla\phi, A^2_0=(1-\cos\theta^0_i)\partial_{\tau}\phi(1-\cos\theta^0_j)\partial_{\tau}\phi$ in conforming with Eq.(\ref{gaugefield}) and symmetrization in going from discrete lattice theory to continuum theory.The geometry of the spin configuration with the 12 sublattice spins is more tedious to work with but the $\Delta\phi^0_{ij},\theta^0_i,\theta^0_j$ can be determined from the projection of a cuboctahedron on the $xy$ plane and the $xz$ plane, from which we obtain
\begin{widetext}
\[
\fl H= H_0+\alpha_{nn} J_1N^c_{nn}\frac{S^2}{2\mathcal{A}}\int d^2\mathbf{r}({\mathbf{A}^{\parallel}}^2(\mathbf{r})+A^2_0(\mathbf{r}))+3\alpha_{nnn} J_2\frac{S^2}{2\mathcal{A}}N^c_{nnn}\int d^2\mathbf{r}({\mathbf{A}^{\parallel}}^2(\mathbf{r})+A^2_0(\mathbf{r}))
\]
\begin{equation}\label{gaugefieldHamiltoniancuboc1result}
+4\alpha_{nnnn} J_3\frac{S^2}{2\mathcal{A}}N^c_{nnnn}\int d^2\mathbf{r}({\mathbf{A}^{\parallel}}^2(\mathbf{r})+A^2_0(\mathbf{r}))
\end{equation}
\end{widetext}
where
\[
\alpha_{nn}=\frac{1}{12}-\frac{1}{6}\left(1-\sqrt{\frac{\sqrt{3}+\sqrt{2}}{\sqrt{3}-\sqrt{2}}}-\sqrt{\frac{\sqrt{3}-\sqrt{2}}{\sqrt{3}+\sqrt{2}}}\right)
\]
\[
\simeq 0.494017,
\]
\[
\alpha_{nnn}=-\frac{1}{6}\left[\frac{1}{2}+\frac{2}{\sqrt{3}}-\frac{1}{6}\left(\frac{\sqrt{3}+\sqrt{2}}{\sqrt{3}-\sqrt{2}}+\frac{\sqrt{3}-\sqrt{2}}{\sqrt{3}+\sqrt{2}}\right)\right]
\]
\[
\simeq 0.00199435
\]
\begin{equation}\label{weightcuboc}
\alpha_{nnnn}=\frac{1}{2}
\end{equation}
and $N^c_{nn}=4,N^c_{nnn}=4,N^c_{nnnn}=2$ are the coordination numbers of the first, second, and third (across the diagonal of the hexagon) neighbor links, respectively.The positiveness of the coefficients $\alpha_{nn},\alpha_{nnn},\alpha_{nnnn}$ is a reassuring justification of the validity of our low energy effective gauge field Hamiltonian to describe stable quantum fluctuations around the ordered state configuration.

Very surprisingly, we note from Eq.(\ref{weightcuboc}) that the third neighbor's weight factor $\alpha_{nnnn}$ is of the same order of magnitude as the first neighbor's weight factor $\alpha_{nn}$ but is much larger than that of second neighbor $\alpha_{nnn}$.This simply means the quantum fluctuations induced by the third neighbor spin exchange interaction is, contrary to naive expectation, nearly as equally strong as those induced by the first neighbor and even much larger than those induced by the second neighbor.As a comparison, we computed the low-energy gauge field Hamiltonian of the $J_1-J_2$ kagom\'{e} Heisenberg model of \cite{Domenge} and found that the weights of the gauge field quadratic terms are of the same order of magnitude between the first and second neighbor contributions in both the ferromagnet and cuboc states (Note that despite having also 12 sublattices, the cuboc state in the $J_1-J_2$ model has different spin configuration than the cuboc1 state in the $J_1-J_2-J_3$ model \cite{Messio}\cite{Domenge}).More precisely, $\alpha_{nn}=\alpha_{nnn}=1/2$ in ferromagnet state and $\alpha_{nn}=-(1/\sqrt{2})\mathrm{tan}[\pi/8]\approx -0.292893,-(1/\sqrt{2})\mathrm{cotan}[\pi/8]\approx -1.70711, \alpha_{nnn}=1$ for the cuboc state.We thus however note that the weights are opposite in overall sign; negative in first neighbor and positive in second neighbor in the cuboc state, which means if the first neighbor spin exchange is much larger than that of the second neighbor, we still need a third neighbor to stabilize the theory working in a Gaussian theory or else we have to include quartic terms.The result reflected in Eqs.(\ref{gaugefieldHamiltoniancuboc1result}) and (\ref{weightcuboc}) clearly implies that the third neighbor term in kagom\'{e} antiferromagnet contributes significantly to inducing quantum fluctuations that will potentially destroy ordered ground state, despite coming from further neighbor interaction.This is the main result of this section.

\section{Chiral Order Parameter Theory for Chiral Ordered Spin State in Kagom\'{e} Antiferromagnet}
An important first step to study chiral spin liquid state is to make a proper definition of it.We define chiral spin liquid to be a quantum disordered spin state at $T=0$ with \textit{no long range spin order (or correlation)} but yet has \textit{long-range chiral order}.We will use chiral order parameter theory first put forward in \cite{Baskaran} to prove that chiral spin state should exist in kagom\'{e} antiferromagnet with next-nearest neighbor (second neighbor) and next-next nearest neighbor spin exchange interactions below critical temperature characterizing the chiral phase transition.The main idea is that with only nearest-neighbor spin exchange interaction, the effective free energy of the spin system has novel local $\mathbb{Z}_2$ symmetry, defined in terms of chiral variable $m_{ijk}$ which is a pseudoscalar quantity related to the scalar chirality $\chi_{ijk}$ and satisfies the following property
\begin{equation}\label{chiralvariableconstraint}
m_{ijk}(\tau)=-m_{ikj}(\tau),m_{ijk}(\tau+\beta)=m_{ijk}(\tau)
\end{equation}
where $i,j,k$ label the sites of any triangle of the kagom\'{e} lattice, corresponding to the three spins living in the triangle and $\tau$ is imaginary time.When one adds second neighbor spin exchange on triangular antiferromagnet, it was found in \cite{Baskaran} that the effective free energy gains chiral symmetry-breaking term, which reduces the local $\mathbb{Z}_2$ symmetry above to the global $\mathbb{Z}_2$ symmetry, that is, the Ising symmetry.Then, since 2d Ising model has finite temperature phase transition, the triangular antiferromagnet with first two neighbor spin exchanges also has finite temperature phase transition, corresponding to chiral phase transition to chiral ordered state \cite{Baskaran}.

Due to the Elitzur theorem \cite{Elitzur}, which asserts that only local gauge invariant operator can have nonzero expectation value, this local $\mathbb{Z}_2$ symmetry, which is not gauge invariant since it corresponds to $\pm 1$ flux, which does not correspond to the gauge invariant spin singlet state, thus cannot be broken at any finite temperature.In other words, we cannot have chiral symmetry breaking at any finite temperature in spin system made up of triangle as the unit cell, with only nearest-neighbor spin exchanges.However, once we add next-nearest neighbor (second neighbor) and next-next-nearest neighbor (third neighbor) spin exchange interactions on kagom\'{e} antiferromagnet, one will obtain Ising-like terms with global rather than local $\mathbb{Z}_2$ symmetry.
 
Following the analysis of Baskaran \cite{Baskaran}, one can compute the effective free energy functional of the system by considering the partition function of the system derived from the Hamiltonian.We rewrite the Hamiltonian defined on kagom\'{e} lattice by making use of the Wen-Wilczek-Zee \cite{WenWilczekZee} identity for the scalar spin chirality $\chi_{ijk}$
\begin{equation}
\hat{\chi}^2_{ijk}=\frac{15}{64}-\frac{1}{64}\left(\mathbf{S}^2_i+\mathbf{S}^2_j+\mathbf{S}^2_k\right)^2
\end{equation}
which re-expresses the Hamiltonian Eq.(\ref{HeisenbergAFMkagome}) as 
\begin{widetext}
\begin{equation}\label{Hamiltonianchiral}
\fl H=-8J_1\sum_{ijk\in\bigtriangleup}(\hat{\chi}^{\bigtriangleup}_{ijk})^2-8J_1\sum_{ijk\in\bigtriangledown}(\hat{\chi}^{\bigtriangledown}_{ijk})^2+\frac{15}{4}-\frac{9}{4}NJ_1+J_2\sum_{ij\in nnn}\mathbf{S}_i\cdot\mathbf{S}_j+J_3\sum_{ij\in nnnn}\mathbf{S}_i\cdot\mathbf{S}_j
\end{equation}
\end{widetext}
where we have introduced the scalar spin chirality for the up and down triangles separately for the following reasons.First, the nearest-neighbor links of the kagom\'{e} lattice has a bow tie unit cell consisting of up and down triangles sharing a corner site.Second, the separation into up and down triangles scalar spin chirality will be useful once we consider further neighbor spin exchange interactions, where different distance neighbor connects different combinations of up and down triangles.The partition function is given by
\begin{widetext}
\begin{equation}\label{partitionfunctionoriginal}
\fl Z(\beta)=\mathrm{Tr}e^{-\beta H}=\mathrm{Tr}\left[\mathcal{T}\left(e^{8J_1\sum_{ijk\in\bigtriangleup}\int^{\beta}_0d\tau\hat{\chi}^2_{ijk}+8J_1\sum_{ijk\in\bigtriangledown}\int^{\beta}_0d\tau\hat{\chi}^2_{ijk}-J_2\sum_{ij\in nnn}\int^{\beta}_0d\tau\mathbf{S}_i\cdot\mathbf{S}_j-J_3\sum_{ij\in nnnn}\int^{\beta}_0d\tau\mathbf{S}_i\cdot\mathbf{S}_j}\right)\right]
\end{equation}
\end{widetext}
where $\beta=1/T$ and $\mathcal{T}$ is time-ordering operator that allows us to treat the operators in the argument of the trace $\mathrm{Tr}$ as if they were commuting and we have absorbed the constant energy factor $E_c=\frac{15}{4}-\frac{9}{4}NJ_1$ in Eq.(\ref{Hamiltonianchiral}) into an overall constant of the partition function.Following G.Baskaran \cite{Baskaran}, we first consider nearest-neighbor exchange only model by setting $J_2=J_3=0$.In this case, the spins at the triangles up and down (apart from that at the shared corner between up and down triangles in a bow tie plaquette) are completely decoupled and can be treated independently.This way, we can then perform Hubbard-Stratonovich transformation in terms of the chiral variables $m^{\bigtriangleup}_{ijk},m^{\bigtriangledown}_{ijk}$ satisfying Eq.(\ref{chiralvariableconstraint}) associated with the up and down triangles, respectively and obtain
\begin{widetext}
\begin{equation}
\fl Z(\beta)=\int\mathcal{D}m^{\bigtriangleup}\mathcal{D}m^{\bigtriangledown}\mathrm{Tr}\left[\mathcal{T}\left(e^{-8J_1\sum_{ijk\in\bigtriangleup}\int^{\beta}_0d\tau (m^{\bigtriangleup}_{ijk})^2+16J_1\sum_{ijk\in\bigtriangleup}\int^{\beta}_0d\tau \chi^{\bigtriangleup}_{ijk}m^{\bigtriangleup}_{ijk}+(\bigtriangleup\rightarrow\bigtriangledown)}\right)\right]
\end{equation}
\end{widetext}
equivalent to the partition function 
\begin{equation}
Z(\beta)=\int\mathcal{D}m^{\bigtriangleup}\mathcal{D}m^{\bigtriangledown}e^{-\beta F[m^{\bigtriangleup},m^{\bigtriangledown}]}
\end{equation}
The resulting free energy is obtained by expanding the right hand side of Eq.(\ref{partitionfunctionoriginal}) in powers of the chiral variables $m^{\bigtriangleup},m^{\bigtriangledown}$ from which we obtain
\begin{widetext}
\[
\fl \beta F[m^{\bigtriangleup},m^{\bigtriangledown}]=8J_1\sum\int^{\beta}_0d\tau(m^{\bigtriangleup(\bigtriangledown)}_{ijk}(\tau))^2-N \mathrm{log}2-\beta\frac{(16J_1)^2}{2!}(\mathrm{Tr}\chi^2)\sum\int^{\beta}_0 d\tau (m^{\bigtriangleup(\bigtriangledown)}_{ijk}(\tau))^2
\]
\[
\fl +\beta^3\frac{(16J_1)^4}{4!}\mathrm{Tr}(\chi^4)\sum\int^{\beta}_0 (m^{\bigtriangleup(\bigtriangledown)}_{ijk}(\tau))^4d\tau-\beta^2\frac{(16J_1)^4}{(2!)^2}\sum \left(\mathrm{Tr}(\chi^2)\right)^2\int^{\beta}_0 d\tau\int^{\beta}_0d\tau'(m^{\bigtriangleup}_{ijk}(\tau))^2(m^{\bigtriangledown}_{jlm}(\tau'))^2
\]
\begin{equation}\label{effectivefreenergy}
\fl +\frac{(16J_1)^4}{4!}\sum\int^{\beta}_0d\tau_1\int^{\beta}_0d\tau_2\int^{\beta}_0d\tau_3\int^{\beta}_0d\tau_4\mathrm{Tr}\left(\mathcal{T}\left[\chi\chi\chi\chi\right]\right)m^{\bigtriangleup}_{ijk}(\tau_1)m^{\bigtriangledown}_{mjn}(\tau_2)m^{\bigtriangleup}_{ijk}(\tau_3)m^{\bigtriangledown}_{mjn}(\tau_4)
\end{equation}
\end{widetext}
The above free energy has local $\mathbb{Z}_2$ symmetry; it is invariant under $m^{\bigtriangleup(\bigtriangledown)}_{ijk}\rightarrow\sigma_{ijk}m^{\bigtriangleup(\bigtriangledown)}_{ijk}$ where $\sigma_{ijk}=\pm 1$.This corresponds to the $\mathbb{Z}_2$ flux of the vortex associated with chiral configuration of the spins on the triangle.This $\mathbb{Z}_2$ degree of freedom is also reflected by the entropy term $-N \mathrm{log}2$.The last term contains nontrivial quartet chiral variable interaction with (imaginary) time-dependent interaction which describes nonlocal interaction in imaginary time, involving pairs of corner-sharing up and down triangles.This high temperature expansion is convergent in both high temperature and low temperature regimes and thus gives rise to no finite temperature chiral symmetry breaking transition \cite{Baskaran}.

The above local $\mathbb{Z}_2$ symmetry however is spoiled once we add further neighbor spin exchange interactions.Recovering $J_2,J_3\neq 0$ in Eq.(\ref{partitionfunctionoriginal}), chiral symmetry-breaking terms immediately appear.Considering the kagom\'{e} lattice geometry and the Heisenberg Hamiltonian Eq.(\ref{HeisenbergAFMkagome}), one can re-compute the free energy with nonzero second and third neighbor spin exchange interactions and obtain
\begin{widetext}
\[
\fl \beta F_{CSB}=-d_1J^2_1 \sum_{\bigtriangleup\bigtriangledown}\int^{\beta}_0\int^{\beta}_0 d\tau_1d\tau_2a^{\bigtriangleup\bigtriangledown}(\tau_1,\tau_2)m^{\bigtriangleup}_{ijk}(\tau_1)m^{\bigtriangledown}_{jlm}(\tau_2)
\]
\begin{equation}\label{CSBterm}
-d_2J^2_1 \sum_{\bigtriangleup\bigtriangleup(\bigtriangledown\bigtriangledown)}\int^{\beta}_0\int^{\beta}_0 d\tau_1d\tau_2 a^{\frac{\bigtriangleup\bigtriangleup}{\bigtriangledown\bigtriangledown}}(\tau_1,\tau_2)m^{\bigtriangleup(\bigtriangledown)}_{ijk}(\tau_1)m^{\bigtriangleup(\bigtriangledown)}_{mln}(\tau_2)
\end{equation}
\end{widetext}
where $d_1, d_2$ are positive numerical factors of the same order of magnitude.The important point is the dependence of the above free energy on the correlation function $a^{\bigtriangleup\bigtriangledown}(\tau_1,\tau_2)$ and $a^{\bigtriangleup\bigtriangleup}(\tau_1,\tau_2)$ given by
\begin{widetext}
\[
\fl a^{\bigtriangleup\bigtriangledown}(\tau_1,\tau_2)=\frac{\mathrm{Tr}\left[\mathcal{T}\left[e^{\left(-\int^{\beta}_0\left(J_2\sum_{ij\in nnn}\mathbf{S}_i(\tau)\cdot\mathbf{S}_j(\tau)+J_3\sum_{ij\in nnnn}\mathbf{S}_i(\tau)\cdot\mathbf{S}_j(\tau)\right)d\tau\right)}\chi^{\bigtriangleup}_{ijk}(\tau_1)\chi^{\bigtriangledown}_{jlm}(\tau_2)\right]\right]}{\mathrm{Tr}\left[\exp\left(-\beta\left(J_2\sum_{ij\in nnn}\mathbf{S}_i\cdot\mathbf{S}_j+J_3\sum_{ij\in nnnn}\mathbf{S}_i\cdot\mathbf{S}_j\right)\right)\right]}
\]
\begin{equation}\label{chiralitycorrelationfunction}
\fl a^{\frac{\bigtriangleup\bigtriangleup}{\bigtriangledown\bigtriangledown}}(\tau_1,\tau_2)=\frac{\mathrm{Tr}\left[\mathcal{T}\left[e^{\left(-\int^{\beta}_0\left(J_2\sum_{ij\in nnn}\mathbf{S}_i(\tau)\cdot\mathbf{S}_j(\tau)+J_3\sum_{ij\in nnnn}\mathbf{S}_i(\tau)\cdot\mathbf{S}_j(\tau)\right)d\tau\right)}\chi^{\frac{\bigtriangleup}{\bigtriangledown}}_{ijk}(\tau_1)\chi^{\frac{\bigtriangleup}{\bigtriangledown}}_{lmn}(\tau_2)\right]\right]}{\mathrm{Tr}\left[\exp\left(-\beta\left(J_2\sum_{ij\in nnn}\mathbf{S}_i\cdot\mathbf{S}_j+J_3\sum_{ij\in nnnn}\mathbf{S}_i\cdot\mathbf{S}_j\right)\right)\right]}
\end{equation}
\end{widetext}
An explicit analytical calculation of the above correlation functions using the Heisenberg Hamiltonian is a challenging task.However, we will be interested in the ground state of the kagom\'{e} antiferromagnet (i.e. at T=0) and we are able to make a crucial advance using the following observation.Analytical (e.g. \cite{Sachdev1992}) and numerical works (e.g.\cite{Huse}\cite{Chalker}\cite{Messio}\cite{Domenge}) found that the ground state of kagom\'{e} antiferromagnet is dominated by magnetically ordered states.Therefore, at $T=0$ or very low temperatures, it is justified in the lowest order approximation, to do calculation using a low-energy effective Hamiltonian instead of the original Heisenberg Hamiltonian.We basically have to analyze the behavior of $\chi_{ijk}$ defined at two different imaginary times $\tau_1$ and $\tau_2$.Since numerics concluded that the chiral spin liquid exists in proximity to spin ordered states \cite{Messio}\cite{Lauchli}, we can treat the spin as classical vector and expand it around an energy-minimizing orientation. That is, $\mathbf{S}_i(\tau)=\mathbf{S}^0_i+\delta \mathbf{S}_i(\tau)$, where $\mathbf{S}^0_i$ is fixed, independent of (imaginary) time $\tau$.The scalar spin chirality then becomes
\begin{widetext}
\[
\fl \chi_{ijk}(\tau)=(\mathbf{S}^0_i+\delta \mathbf{S}_i(\tau))\cdot\left((\mathbf{S}^0_j+\delta \mathbf{S}_j(\tau))\times(\mathbf{S}^0_k+\delta \mathbf{S}_k(\tau))\right)
\]
\begin{equation}\label{scalarspinchiralityfluctuation}
=\chi^0_{ijk}+\delta\mathbf{S}_i(\tau)\cdot(\mathbf{S}^0_j\times\mathbf{S}^0_k)+\mathbf{S}^0_i\cdot(\delta\mathbf{S}_j(\tau)\times\mathbf{S}_k)+\mathbf{S}^0_i\cdot(\mathbf{S}^0_j\times\delta\mathbf{S}_k(\tau))
\end{equation}
\end{widetext}
where $\chi^0_{ijk}=\mathbf{S}^0_i\cdot(\mathbf{S}^0_j\times\mathbf{S}^0_k)$.Now, for the ordered states of interest; the $\sqrt{3}\times\sqrt{3}$, $\mathbf{q}=0$, and cuboc1 states, it turns out that the three spins in each elementary triangle are always coplanar.This means that $\chi^0_{ijk}=0$ identically and the leading contribution to the dynamical scalar spin chirality comes from the first order in the spin fluctuations given in Eq.(\ref{scalarspinchiralityfluctuation}), which is nonzero since the spin vectors are free to fluctuate normal to the coplanar plane on which each set of three spins on each triangle live.Such fluctuations are described together by $\delta\phi$ and $\delta\theta$, which we treat on equal footing in the following calculations.

First, we analyze the expression Eq.(\ref{scalarspinchiralityfluctuation}) qualitatively for the two coplanar states; the $\sqrt{3}\times\sqrt{3}$ and $\mathbf{q}=0$ states.The spin configurations for these states are given in \cite{Sachdev1992}.It can be seen for the $\sqrt{3}\times\sqrt{3}$ state, at two bowtie clusters separated by third neighbor distance across the hexagon, the spin configurations are flipped with respect to each other.On the other hand, for the $\mathbf{q}=0$ state, the spin configurations at the two bowties are identical.On the other hand, the spin configuration at the up and down triangles in each bowtie is mirror symmetric (with respect to mirror reflection about $x$ axis) in the $\sqrt{3}\times\sqrt{3}$ state, but is not so in the $\mathbf{q}=0$ state.Now we consider the noncoplanar cuboc1 state.It turns out that in cuboc1 state, the three spins on each elementary triangle are also always coplanar, even though between two different elementary triangles the spins are non-coplanar.As a result, we also have $\chi^0_{ijk}=0$ for the cuboc1 state.It is to be noted that even though the (static) scalar spin chirality vanishes $\chi^0_{ijk}$ for all the three ordered states we are considering, its \textit{fluctuations} do not vanish, and so its correlation function does not vanish either.We use Green's function method within semiclassical approximation to compute the above correlation functions, within continuum formulation and employing harmonic approximation.In this case, the expectation value of the product of two functions is given by
\begin{equation}
\langle \mathcal{T}\left[A(\tau_1)B(\tau_2)\right]\rangle=\frac{\int d\zeta \mathcal{T}\left[e^{-S[\zeta]}A(\tau_1)B(\tau_2)\right]}{\int d\zeta e^{-S[\zeta]}}
\end{equation}
In this case, we have the functions $A$ and $B$ are the scalar spin chirality while the $\zeta$ describes the quantum fluctuation phase variable $\zeta=(\delta\phi,\delta\theta)^T$ with the action given by 
\[
S[\zeta]=\int d\tau \left[\zeta^T(\partial_{\tau}+H[\zeta])\zeta+J(\tau)\delta\theta(\tau)\right]
\]
where the Hamiltonian $H[\zeta]$ is given by the quantum fluctuating part of the effective Hamiltonian in Eq.(\ref{Hamiltonianfluctuation}) and is to be written in terms of Green's function
\begin{equation}
H[\zeta(\tau)]=\int d\mathbf{r}\int d\mathbf{r}'\zeta^T(\mathbf{r},\tau)G^{-1}(\mathbf{r},\mathbf{r}',\tau)\zeta(\mathbf{r}',\tau)
\end{equation}
and we work in the Euclidean space-time (with imaginary time).For the purpose of the computation of correlation function, as is standard in quantum field theory 
\cite{ZinnQFT}\cite{RyderQFT} one has to add a `source term' that couples to the physical field; the polar angle fluctuation field in this case, which gives a $J(\tau)\delta\theta(\tau)$ term.Physically, the source current can be realized by applying a Zeeman magnetic field along $+z$ direction, giving rise to Zeeman term $H_Z=-\mu \mathbf{S}\cdot\mathbf{B}$, which gives $J(\tau)=\mu S B(\tau) \sin\theta^0$ for small fluctuations around the ground state polarization polar angle $\theta^0$ in the Hamiltonian.This source current $J$ is to be taken to zero at the end of the calculation \cite{ZinnQFT}\cite{RyderQFT}.

Now, for the scalar spin-chirality correlation function of interest, we have
\begin{equation}
\langle \mathcal{T}\left[\chi^{\bigtriangleup}_{ijk}(\tau_1)\chi^{\bigtriangledown}_{jlm}(\tau_2)\right]\rangle=\frac{\int d\zeta \mathcal{T}\left[e^{-S[\zeta]}\chi^{\bigtriangleup}_{ijk}(\tau_1)\chi^{\bigtriangledown}_{jlm}(\tau_2)\right]}{\int d\zeta e^{-S[\zeta]}}|_{J=0}
\end{equation}   
connected by second neighbor spin exchange and similarly for $\langle \mathcal{T}\left[\chi^{\frac{\bigtriangleup}{\bigtriangledown}}_{ijk}(\tau_1)\chi^{\frac{\bigtriangleup}{\bigtriangledown}}_{mln}(\tau_2)\right]\rangle$ connected by third neighbor spin exchange.For the three ordered states of our interest, $\langle \mathcal{T}\left[\chi^{\bigtriangleup}_{ijk}(\tau_1)\chi^{\bigtriangledown}_{jlm}(\tau_2)\right]\rangle=\langle \mathcal{T}\left[\delta\chi^{\bigtriangleup}_{ijk}(\tau_1)\delta\chi^{\bigtriangledown}_{jlm}(\tau_2)\right]\rangle$.The details of the derivation of these correlation functions are given in the Appendix A.The main result is as follows.As derived in the Appendix A, the correlation function is found to be

\begin{equation}\label{chiralitycorrfunctiontemporal}
\fl \langle \mathcal{T}\left[\chi_{ijk}(\tau_1)\chi_{mln}(\tau_2)\right]\rangle=S^6f_{ijk}(\phi^0,\theta^0)f_{mln}(\phi^0,\theta^0)\frac{1}{2}\int_{\mathbf{x}_1}\int_{\mathbf{x}_2}\frac{1}{\partial_{\tau}+\tilde{g}^{-1}_{\mathbf{x}_1\mathbf{x}_2}(\partial^2_{\tau}+\nabla^2)}
\end{equation}
%\end{widetext}
where $f_{ijk}(\phi^0,\theta^0)$ is what we will refer to as a scalar spin chirality `form factor' with its detailed expression given in Appendix A, $\tau=\tau_2-\tau_1$ and 
\begin{widetext} 
\begin{equation}\label{gfunction}
\fl \tilde{g}^{-1}_{\mathbf{x}_1\mathbf{x}_2}=\frac{1}{2}S^2J_{ij}\delta(|\Delta\mathbf{x}_{12}|-|\Delta\mathbf{r}_{ij}|)\left[\cos\Delta\phi^0_{\mathbf{x}_1\mathbf{x}_2}\sin\theta^0_{\mathbf{x}_1}\sin\theta^0_{\mathbf{x}_2} +\cos\theta^0_{\mathbf{x}_1}\cos\theta^0_{\mathbf{x}_2}\right]
\end{equation}
\end{widetext}
with $\Delta\mathbf{x}_{12}=\mathbf{x}_2-\mathbf{x}_1$.For this calculation, according to Eq.(\ref{chiralitycorrelationfunction}), we have $|\Delta\mathbf{r}_{nnn}|=\sqrt{3},|\Delta\mathbf{r}_{nnnn}|=2$ for the second and third neighbor spin exchanges, respectively.

We evaluate the above Eqs.(\ref{chiralitycorrfunctiontemporal}-\ref{gfunction}) for the coplanar $\mathbf{q}=0$ and $\sqrt{3}\times\sqrt{3}$ states where $\Delta\phi^0_{ij\in nnn}=2\pi/3,\Delta\phi^0_{ij\in nnnn}=0$ and $\Delta\phi^0_{ij\in nnnn}=0,\Delta\phi^0_{ij\in nnnn}=2\pi/3$ respectively and also for the cuboc1 state.The details of the resulting expressions and their analysis are given in Appendix A.The important outcome of this analysis is that, for the $\mathbf{q}=0$ and $\sqrt{3}\times\sqrt{3}$ states, the $\tilde{g}^{-1}_{\mathbf{x}_1\mathbf{x}_2}|_{\mathbf{q}=0}$ has both negative-valued and positive-valued components, coming from  $J_2$ and $J_3$ contributions respectively, whereas $\tilde{g}^{-1}_{\mathbf{x}_1\mathbf{x}_2}|_{\sqrt{3}\times\sqrt{3}}$ is always negative in sign, regardless of the relative magnitude of $|J_2|$ and $|J_3|$.On the other hand, very interestingly, for the noncoplanar cuboc1 state, the second and third neighbor contributions are of precisely the same weight but opposite in sign but what is important is it does have positive component, which comes from second neighbor contribution at the border between the cuboc1 and $\mathbf{q}=0$ state.As will be shown soon below, positive value for $\tilde{g}^{-1}_{\mathbf{x}_1\mathbf{x}_2}$ gives rise to strong chirality effect whereas negative value for $\tilde{g}^{-1}_{\mathbf{x}_1\mathbf{x}_2}$ gives rise to weak chirality effect.This will have crucial consequence as we will demonstrate in what follows.

Performing Fourier transform on Eq.(\ref{chiralitycorrfunctiontemporal}), we obtain
%\begin{widetext}
\begin{equation}\label{chiralitycorrfunctiontemporalfinal}
\langle \mathcal{T}\left[\chi_{ijk}(\tau_1)\chi_{mln}(\tau_2)\right]\rangle=S^6f_{ijk}(\phi^0,\theta^0)f_{mln}(\phi^0,\theta^0)\frac{1}{2}\int_{\mathbf{x}_1}\int_{\mathbf{x}_2}F_{\mathbf{x}_1\mathbf{x}_2}(\tau_2-\tau_1)
\end{equation}
%\end{widetext}
where 
%\begin{widetext}
\begin{equation}\label{temporalcorrelation}
\fl F_{\mathbf{x}_1\mathbf{x}_2}(\tau)=-\frac{\omega_0e^{\omega_0\tau} \sqrt{2\pi}}{\omega_{\mathbf{k}}} \left(e^{ -\omega_{\mathbf{k}} \tau}\Theta[\tau \mathcal{S}[\omega_{\mathbf{k}} - \omega_0]] \mathcal{S}[\omega_{\mathbf{k}} - \omega_0]+e^{ \omega_{\mathbf{k}} \tau}
        \Theta[-\tau \mathcal{S}[\omega_{\mathbf{k}} + \omega_0]] \mathcal{S}[\omega_{\mathbf{k}} + \omega_0]\right)
\end{equation}
%\end{widetext}
where $\mathcal{S}(\cdots)$ refers to the sign of $(\cdots)$ and 
\begin{equation}\label{coefficientsinF}
\omega_0=\frac{1}{2\tilde{g}^{-1}_{\mathbf{x}_1\mathbf{x}_2}},\omega^2_{\mathbf{k}}=\frac{1}{4(\tilde{g}^{-1}_{\mathbf{x}_1\mathbf{x}_2})^2}-\nabla^2 =\mathrm{I.F.T}\left[\frac{1}{4}\mathcal{G}_{\mathbf{k}}+\mathbf{k}^2\right]
\end{equation}
with $\Theta[z]$ the Heaviside theta function; $\Theta[z]=1$ for $z\geq 0$ and zero otherwise and we have symbolically labeled the (spatial) Fourier transform by $\mathrm{F.T.}$(e.g. $\mathcal{G}_{\mathbf{k}}=\mathrm{F.T.}\left[1/(\tilde{g}^{-1}_{\mathbf{x}_1\mathbf{x}_2})^2\right]$) and its inverse by $\mathrm{I.F.T}$.

We note the interesting temporal correlation with its nontrivial dependence on the imaginary time $\tau$.Typical profiles of the above temporal correlation function is illustrated in Fig. 2 for $T=0$.We note that once we fix $\omega_{\mathbf{k}}>0$, for $\omega_0>0$ we obtain negative-valued $F_{\mathbf{k}}(\tau)$, as shown in Fig.2a).This result is applicable to the $\mathbf{q}=0$ for its third neighbor contribution and also to the second neighbor contribution of the cuboc1 state at its border with the $\mathbf{q}=0$ state, where $\tilde{g}^{-1}_{\mathbf{x}_1\mathbf{x}_2}|_{\mathbf{q}=0}>0$ so that $\omega_0=1/(2\tilde{g}^{-1}_{\mathbf{x}_1\mathbf{x}_2}|_{\mathbf{q}=0})>0$ too.On the other hand, for $\omega_0<0$, we obtain positive-valued $F_{\mathbf{k}}(\tau)$, as shown in Fig. 2b).We note that in terms of order of magnitude, the negative-valued $F_{\mathbf{k}}(\tau \approx 0)$ is of the same order of magnitude as that of the positive-valued $F_{\mathbf{k}}(\tau\approx 0)$ but the former decays far slower than that of the latter.As one goes from $\tau=0$ to $\tau=\infty$, the temperature decreases from $T=\infty$ to $T=0$.We note that for most ranges of temperatures, the negative-valued temporal correlation dominates over the positive-valued one in magnitude.These results illustrate low momentum $|\mathbf{k}|\rightarrow 0$ contribution to the temporal correlation according to Eq.(\ref{coefficientsinF}).We found that the large momentum $|\mathbf{k}|\rightarrow \infty$ contribution is strongly suppressed with increasing $|\mathbf{k}|$ and thus do not affect the conclusion deduced from small $|\mathbf{k}|$. 
\begin{figure}
\begin{center}
\includegraphics[scale=0.075]{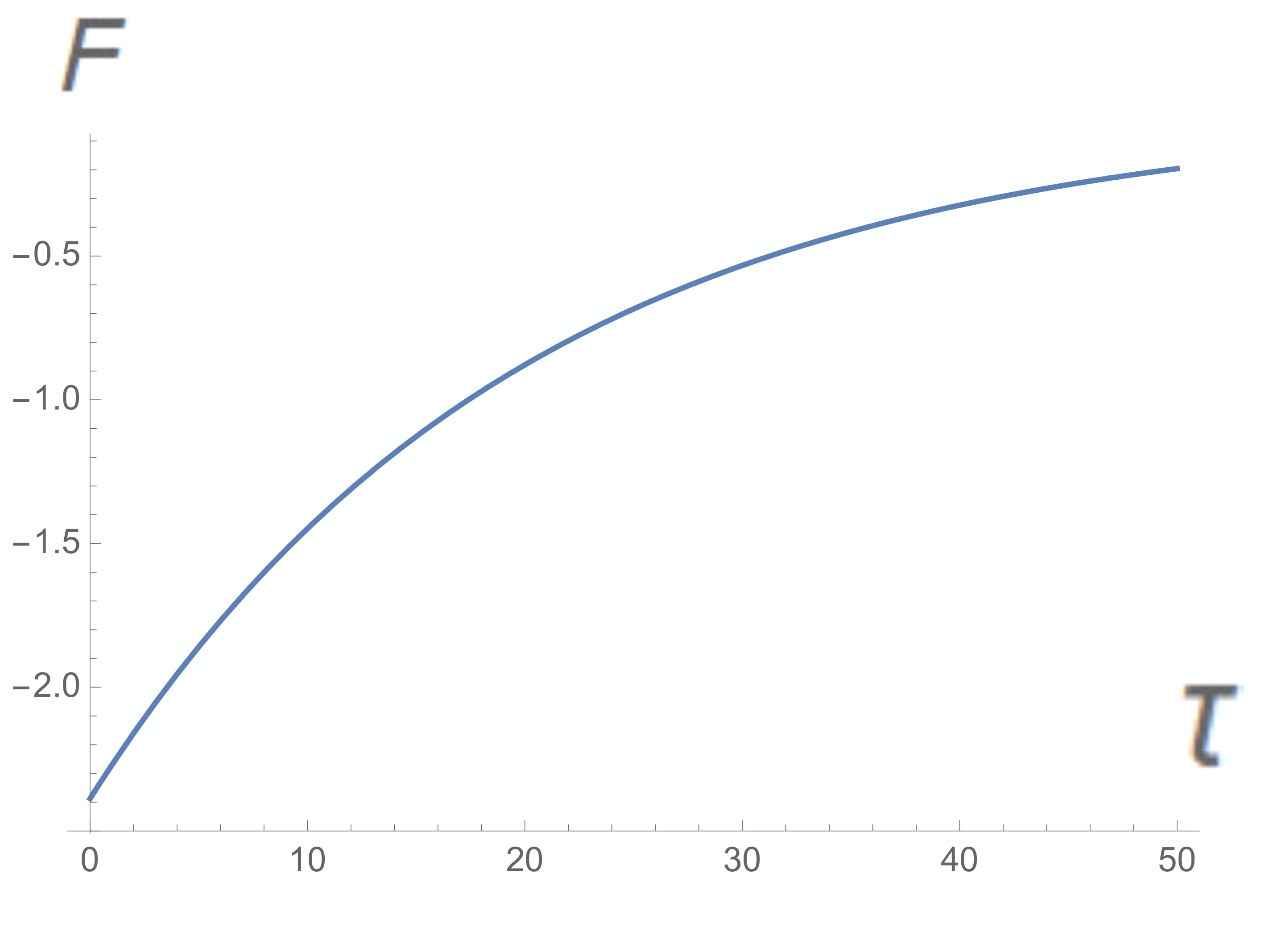}
 \includegraphics[scale=0.075]{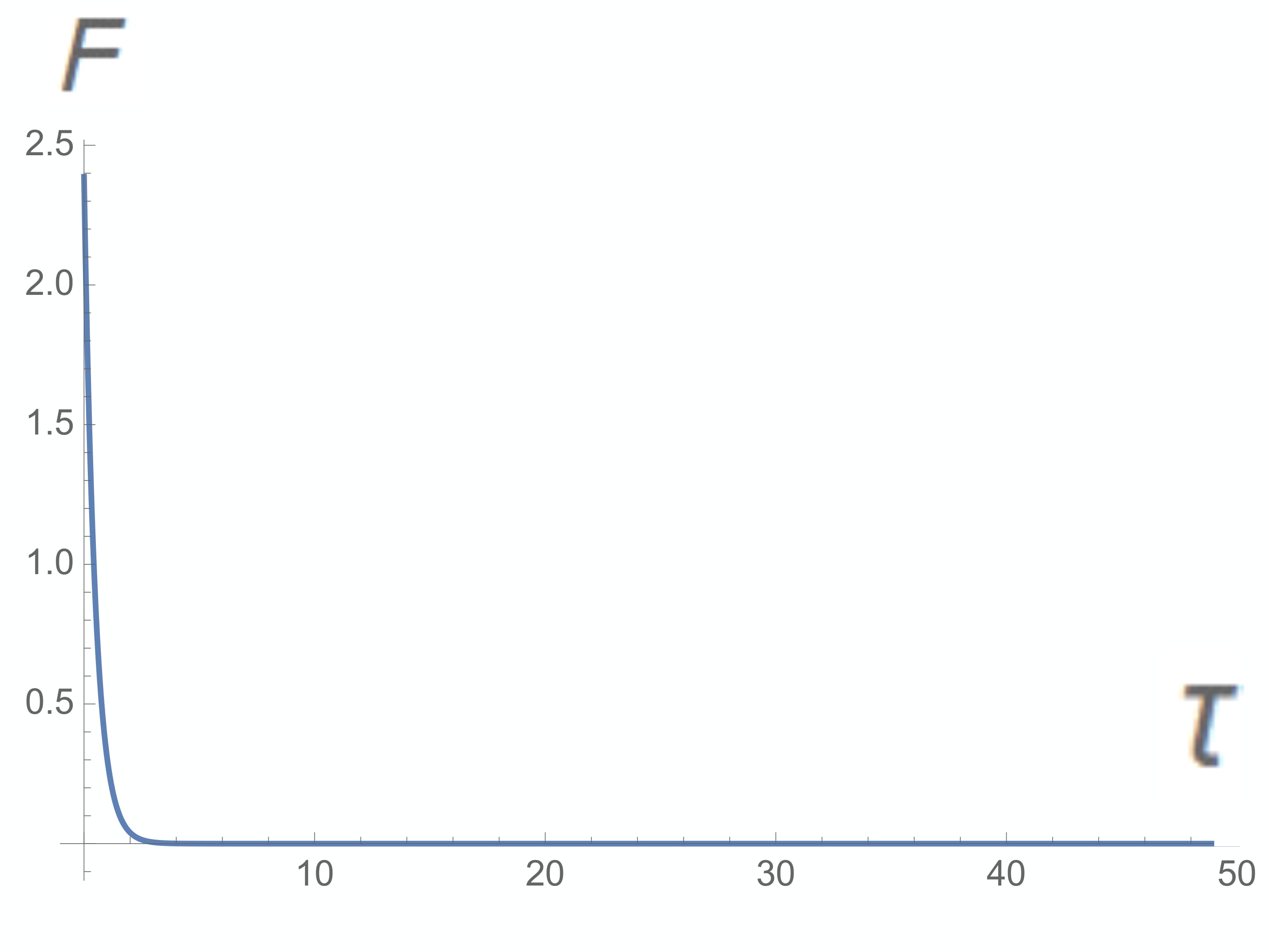}
 \label{fig:Temporalcorrfunction}
 \caption{
 Typical profiles of the temporal correlation function $F_{\mathbf{k}}(\tau)=\mathrm{F.T.}[F_{\mathbf{x}_1\mathbf{x}_2}(\tau)]$, reflecting exponentially-decaying behavior (in magnitude) where we have set a)$\omega_{\mathbf{k}}=1.05,\omega_0=1.0$b)$\omega_{\mathbf{k}}=1.05,\omega_0=-1.0$, these values being chosen to represent the contribution of the low energy long distance physics $|\mathbf{k}|\rightarrow 0$.The profile in a) dominates over that in b) in magnitude $|F_{\mathbf{k}}(\tau)|$ for most of $\tau$.We found that the profile a) appears only in the cuboc1 and $\mathbf{q}=0$ states, but not in $\sqrt{3}\times\sqrt{3}$ state.}
 \end{center}
 \end{figure}

Note that according to Eq.(\ref{chiralitycorrelationfunction}), only the second and third neighbor spin exchanges $J_2,J_3$ contribute to the above result.Eq.(\ref{chiralitycorrfunctiontemporalfinal}) constitutes a key result of this work.Now, according the result shown in Fig.2, the negative-valued $F_{\mathbf{k}}(\tau)$ is much larger in magnitude than that of the positive-valued $F_{\mathbf{k}}(\tau)$ for most temperature ranges.According to Eqs.(\ref{gfactorq=0})-(\ref{gfactorcuboc1}), only third neighbor exchange in the $\mathbf{q}=0$ state consistently gives rise to negative-valued temporal correlator in addition to the one that comes from the second neighbor contribution in the cuboc1 state at its border with the $\mathbf{q}=0$ state, while the rest gives rise to recessive $F_{\mathbf{k}}(\tau)>0$.It is to be noted that the strength of the chiral symmetry-breaking effect is proportional to this magnitude of $|F_{\mathbf{k}}(\tau)|$ in Eq.(\ref{CSBterm}).One thus obtains a significant and observable chiral symmetry-breaking effect only between the cuboc1 and $\mathbf{q}=0$ states in the $J_1-J_2-J_3$ model.This result immediately explains why numerics found chiral spin liquid in the ground state $T=0$ phase diagram only after they included third neighbor spin exchange \cite{Messio}\cite{DNShengSciRepFQHEkagome}\cite{Lauchli}.As comparison, in $J_1-J_2$ model (with $J_1<0$) in kagom\'{e} lattice, one found ferromagnetic and cuboc ground states at $T=0$, with finite temperature chiral ordered state and disordered state separated by finite temperature phase transition between them that ends at $T=0$ at $J_2/|J_1|=1/3$ \cite{Domenge}.In this case, the elementary triangle has nonzero $\pm 1$ scalar spin chirality in the cuboc state at $T=0$ and also in the chiral ordered state at the finite temperature regime above it, compared to the cuboc1 state of the $J_1-J_2-J_3$ kagom\'{e} we are considering in this work where the scalar spin chirality identically vanishes at each elementary triangle since the three spins are always coplanar there.Interestingly, no chiral spin liquid was found  \cite{Domenge}.

Our theory explains this by noting that for this $J_1-J_2$ model, we will get $\tilde{g}^{-1}_{\mathbf{x}_1\mathbf{x}_2\in nnn}=J_2S^2/2>0$ for ferromagnet ground state but $\tilde{g}^{-1}_{\mathbf{x}_1\mathbf{x}_2\in nnn}=-J_2S^2/4<0$ for the cuboc ground state \cite{Domenge}, giving rise to the expected disordered state and chiral ordered state at finite but low temperatures, noting that $J_1<0$ (ferromagnetic)\cite{Domenge}.However, the mismatch in sign of $\tilde{g}^{-1}_{\mathbf{x}_1\mathbf{x}_2\in nnn}$ means one needs further spin exchange interactions until there are contributions of equal sign (that gives rise to strong chiral symmetry-breaking effect) from ordered states on both sides of transition to extend the chiral symmetry breaking effect to $T=0$ to produce chiral spin liquid.The latter is readily realized in $J_1-J_2-J_3$ model.Furthermore, our result naturally explains the numerical conclusion that the chiral spin liquid occurs at the phase transition between the $\mathbf{q}=0$ and cuboc1 states but not between the $\sqrt{3}\times\sqrt{3}$ and cuboc1 states in a $J_1-J_2-J_3$ kagom\'{e} antiferromagnet \cite{Messio}\cite{DNShengSciRepFQHEkagome}\cite{Lauchli}.Again, this is because, according to Eqs.(\ref{gfactorq=0})-(\ref{gfactorcuboc1}) the $\sqrt{3}\times\sqrt{3}$ state always gives rise to $\tilde{g}^{-1}_{\mathbf{x}_1\mathbf{x}_2}<0$ and thus $F_{\mathbf{k}}(\tau)>0$ which is strongly suppressed in magnitude at any finite temperature in the low energy limit.The cuboc1 and $\mathbf{q}=0$ states on the other hand have slowly suppressed $F_{\mathbf{k}}(\tau)<0$ coming from the second and third neighbor contributions respectively, that gives rise to persistent chiral symmetry breaking to $T=0$.

Now, we compute the product of form factors in Eq.(\ref{chiralitycorrfunctiontemporalfinal}).For the $\mathbf{q}=0$ planar ordered state, we obtain from Eq.(\ref{scalarspinchiralityfluctuations}),
\[
f^{\bigtriangleup}_{ijk}f^{\bigtriangledown}_{jmn}=\frac{1}{4}(3+\sqrt{3})^2\simeq 5.59808,
\]
\begin{equation}\label{chiralCFformfactorq=0}
f^{\frac{\bigtriangleup}{\bigtriangledown}}_{ijk}f^{\frac{\bigtriangleup}{\bigtriangledown}}_{mln}=\frac{1}{4}(3+\sqrt{3})^2\simeq 5.59808
\end{equation}
which shows that the chirality-chirality form factor product is not only of the same order of magnitude but also of exactly the same magnitude for the second and third neighbor spin exchange-induced correlations.

For the $\sqrt{3}\times\sqrt{3}$ planar ordered state, we obtain from Eq.(\ref{scalarspinchiralityfluctuations}),
\[
f^{\bigtriangleup}_{ijk}f^{\bigtriangledown}_{jmn}=-\frac{27}{4}=-6.75,
\]
\begin{equation}\label{chiralCFformfactorq=0}
f^{\frac{\bigtriangleup}{\bigtriangledown}}_{ijk}f^{\frac{\bigtriangleup}{\bigtriangledown}}_{mln}=\frac{27}{4}=6.75
\end{equation}
which also shows equal magnitude in the product of chirality factors but which are opposite in sign between that of up-down triangles and up-up or down-down triangles.These opposite signs give rise to Ising interaction of opposite types (antiferromagnetic vs. ferromagnetic) between the contributions of the second and third neighbor exchanges, respectively.However, as we noted earlier, the chiral symmetry breaking free energy terms in this $\sqrt{3}\times\sqrt{3}$ state are strongly suppressed at low temperatures and so this state is no longer relevant in this discussion of chiral symmetry breaking.

For the non-coplanar cuboc1 state, the chirality-chirality form factor product is found to be
\[
f^{\bigtriangleup}_{ijk}f^{\bigtriangledown}_{jmn}=-\frac{3}{4} (1 + 2 \sqrt{3})\simeq -3.34808,
\]
\begin{equation}\label{chiralCFformfactorcuboc1}
f^{\frac{\bigtriangleup}{\bigtriangledown}}_{ijk}f^{\frac{\bigtriangleup}{\bigtriangledown}}_{mln}=-\frac{3}{4} (1 + 2 \sqrt{3})\simeq -3.34808
\end{equation}
These are still of the same order of magnitude as those of coplanar ordered states.However, the crucial difference is in the sign; it is positive for the coplanar $\mathbf{q}=0$ state but is negative for the noncoplanar cuboc1 state.As noted earlier, in Eq.(\ref{chiralitycorrfunctiontemporalfinal}), the dominant contribution occurs when $F_{\mathbf{k}}(\tau)$ is negative-valued, which in combination with Eqs.(\ref{CSBterm}),(\ref{chiralitycorrfunctiontemporalfinal}),(\ref{chiralCFformfactorq=0}),(\ref{chiralCFformfactorcuboc1}), gives rise antiferromagnetic and ferromagnetic Ising interactions for the $\mathbf{q}=0$ and cuboc1 states, respectively.This thus gives antiferromagnetically chiral ordered and ferromagnetically chiral ordered states at low but finite temperatures above the $\mathbf{q}=0$ and cuboc1 ground states respectively.This change in the pattern of finite temperature chiral ordered states is surprising and unprecedented to the best of our knowledge.Interestingly, this dramatic change in the chiral ordering pattern indicates that the $T=0$ transition between the $\mathbf{q}=0$ and cuboc1 states necessitates a complete reorientation of the spins, which eventually corresponds to a quantum disordered spin liquid state. 

The entire discussion above made argument for the occurrence of chiral symmetry breaking at finite temperature and going to the limit $T=0$ and we have shown that third neighbor spin exchange enhances the chiral symmetry breaking effect in kagom\'{e} antiferromagnet to very low temperatures.We want to know, how such chiral symmetry breaking effect manifests itself at $T=0$, in the low-energy effective (continuum) theory of the system.

\section{Emergence of Chern-Simons Effective Theory of Chiral Spin Liquid}

So far, the chiral spin liquid hypothesized to occur in 2d frustrated spin systems has been described by a low energy effective theory which takes the form of the topological Chern-Simons term\cite{WenWilczekZee}\cite{KunYangGirvin}\cite{FradkinSchaposnik}.Other than that this is motivated by the idea that chiral spin liquid is the spin analog of fractional quantum Hall effect \cite{KalmeyerLaughlin}, this is also motivated by the fact that such Chern-Simons term breaks discrete symmetries, such as time-reversal, parity, and chirality, as is expected of a chiral spin liquid state.However, there has never been explicit derivation of such Chern-Simons term applied directly to the putative chiral spin liquid state of kagom\'{e} antiferromagnet in existing studies.Here we will derive and show explicitly that Chern-Simons term arises naturally at the phase transition between two appropriate ordered ground states of 2d frustrated spin systems, applicable to kagom\'{e} antiferromagnet.Consider two ordered states characterized by appropriate phase angle fields $\phi^A(\mathbf{x},t)$ and $\phi^B(\mathbf{x},t)$ which are necessarily dynamical in order to describe quantum fluctuations around the corresponding ordered configurations, corresponding to Goldstone bosons.We propose a scenario for the criticality between the two phases as follows.

%\begin{widetext}
\begin{figure}
\begin{center}
 \includegraphics[scale=0.15]{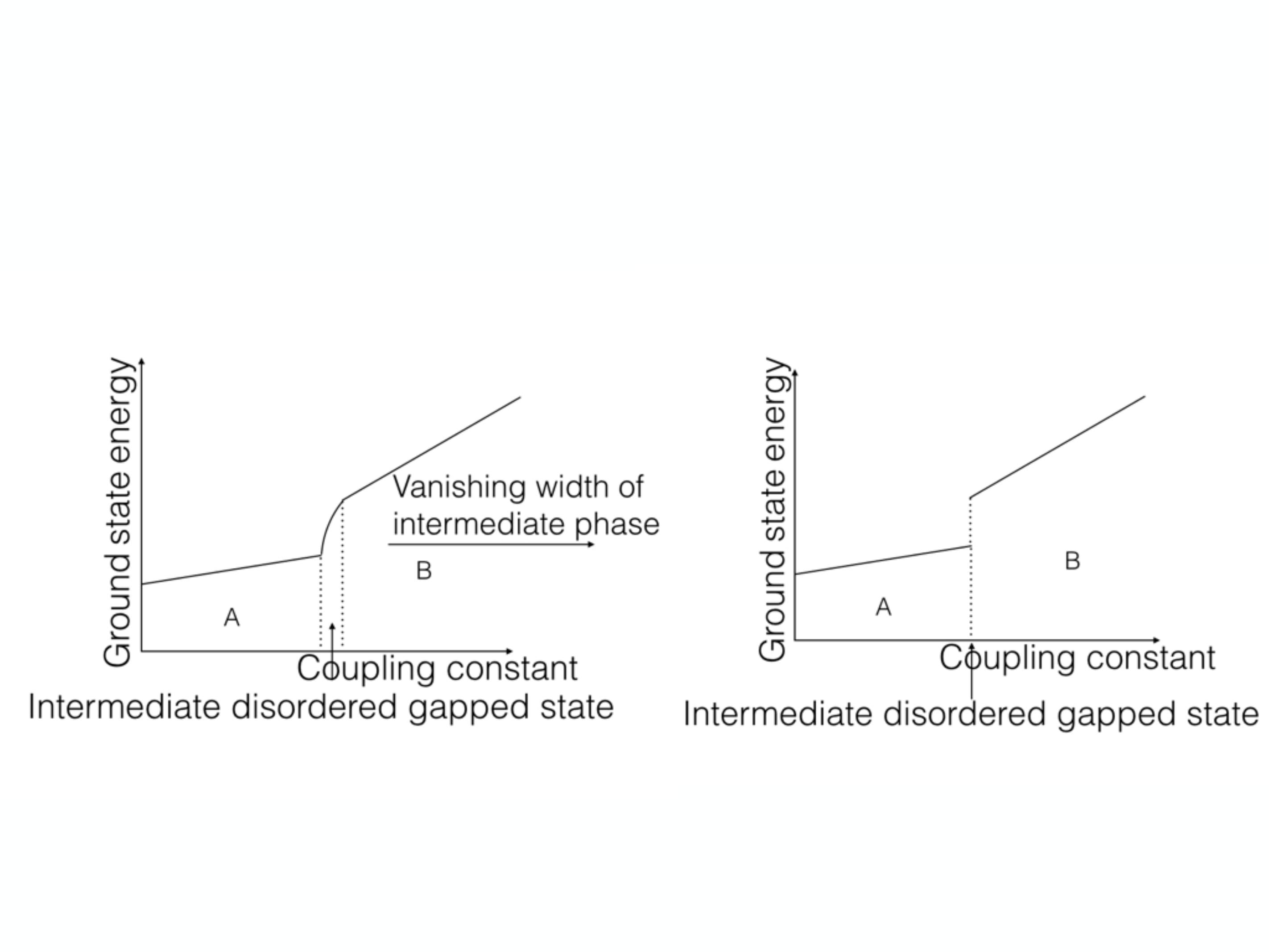}
 \label{fig:Criticality}
 \caption{Our proposed quantum criticality scenario between the ordered ground states in terms of their ground state energy.}
 \end{center}
 \end{figure}
%\end{widetext}

We consider a quantum criticality scenario between two ordered ground states, as illustrated in Fig.3.As one tunes an appropriate coupling constant $J$, one drives the ground state from an ordered state A at one end to another ordered state B at the other end.Since the spin configurations of these two ordered states can be completely different in general and since the tuning of the coupling is continuous, there are two possible scenarios.One is a continuous deformation of the spin configuration from that of ordered state A to that of ordered state B.This is the familiar scenario of quantum phase transition \cite{Sachdev} which, when the width of the intermediate region is finite, gives rise to another ordered state, or, when the width is vanishing, corresponds to a transition line which is continuous despite the different ordered states on the two sides, also known as a deconfined quantum criticality \cite{SenthilScience}\cite{SenthilPRB}.The second scenario is that in going from ordered state A to ordered state B, the spin system is completely disordered first so that it can rearrange itself in going between the two ordered states.This is the scenario that we propose to give rise to a chiral spin liquid state between two ordered states.The question is then how such disordered intermediate state could happen.The properties of such intermediate state are governed by the effective action (Lagrangian) or ground state Hamiltonian (energy) of the spin system.The ground state energy should be continuously dependent on the tuning parameter (coupling constant) in such a way that it connects continuously to the two neighboring ordered states.One can think of the ground state energy difference as function of the tuning parameter in such a way that it is zero at the transition line between the disordered state and the ordered state A and equals the ground state energy difference between the ordered states A and B at the transition line between the disordered state and ordered state B.This ground state energy difference, which is more discernible in the limit of vanishing width of the intermediate disordered state as illustrated in Fig. 3, will then describe the disordered ground state in the interval of the values of the the coupling constant that describe the intermediate state.This ground state energy difference can be obtained from the difference between ground state energies of the two neighboring ordered states.We will compute the action (or free energy) difference that includes this ground state energy difference and the difference in energy of quantum fluctuations around two ordered ground states, outlined according to the following line of reasonings.

Imagine we tune a coupling constant of the spin Hamiltonian in Eq.(\ref{HeisenbergAFMkagome}) slowly $J\rightarrow J+\Delta J$, then the spin configuration changes accordingly.If we consider such tuning across the phase transition from ordered state A represented by order parameter field $\phi^A(\mathbf{r},t)\equiv\phi(J,\mathbf{r},t)$ to ordered state B represented by order parameter field $\phi^B(\mathbf{r},t)\equiv \phi(J+\Delta J,\mathbf{r},t)$, we have at each lattice site $i$
%\begin{widetext}
\[
\fl \phi^B(\mathbf{r}_i,t)=\phi^A(\mathbf{r}_i,t)+\Delta J \frac{\partial\phi^A(\mathbf{r}_i,t)}{\partial J}+\cdots=\phi^A(\mathbf{r}_i,t)+\Delta J \frac{\partial \mathbf{r}_j}{\partial J}\frac{\partial\phi^A(\mathbf{r}_j,t)}{\partial \mathbf{r}_j}|_{\mathbf{r}_i}+\cdots
\]
\begin{equation}\label{expansion}
\fl =\phi^A(\mathbf{r}_i,t)+\Delta r_{j\alpha} \frac{\partial\phi^A(\mathbf{r}_j,t)}{\partial r_{j\alpha}}|_{\mathbf{r}_i}+\cdots=\phi^A(\mathbf{r}_i,t)+\Delta \mathbf{r}_j \cdot \nabla_j\phi^A(\mathbf{r}_j,t)|_{\mathbf{r}_i}+\cdots
\end{equation}
%\end{widetext}
where $j$ denotes all other lattice sites, $\alpha=x,y$ and we have used the standard chain rule and defined
\begin{equation}\label{Jtuneddisplacement}
\Delta \mathbf{r}_j=\frac{\partial \mathbf{r}_j}{\partial J}\Delta J=(\partial J/\partial \mathbf{r}_j)^{-1}\Delta J
\end{equation}
or equivalently, $\Delta \mathbf{r}_j=(\nabla J)^{-1}\Delta J=(\Delta x_{r_j},\Delta y_{r_j})$.It is to be noted that the gradient $\partial J/\partial \mathbf{r}_j$ is defined with respect to vector variable that corresponds at the microscopic lattice level to the relative position vector between two sites $\mathbf{r}_j-\mathbf{r}_{i=0}$ and thus $\partial J/\partial \mathbf{r}$ will be the continuum limit of the gradient of the spin exchange in the relative coordinate with respect to a reference lattice site $\mathbf{r}-\mathbf{r}_{0}$.This can also be seen from the fact that the spin exchange is a function of the relative position vector $J_{ij}=J(\mathbf{r}_j-\mathbf{r}_i)$ and so when we gradient expand $J$ to few lowest orders in the continuum limit, the expansion is with respect to the relative coordinate $(\mathbf{r}_j-\mathbf{r}_i)$.The expansion is not with respect to the link center coordinate $\mathbf{R}_{ij}=(\mathbf{r}_i+\mathbf{r}_j)/2$ because of the above reason as well as the fact that the order parameter is defined at the lattice site $\mathbf{r}_i$ and that the spin exchange interaction occurs between the spins at these lattice sites and not between the link centers $\mathbf{R}_{ij}$.The spatial dependence of the spin exchange $J(\mathbf{r})$ in the continuum limit then corresponds to the distance (and direction)- dependent spin exchange couplings in the original lattice Hamiltonian Eq.(\ref{HeisenbergAFMkagome}), represented by the presence of the first, second, and third neighbor spin exchange coupling constants.It is important to note that with this definition, $\nabla J(\mathbf{r})$ is nonzero even for a microscopic lattice spin model with homogeneous spin exchange $J$'s which thus preserves the lattice translational invariance; the nonzero $\nabla J(\mathbf{r})$ comes from the distance dependence of $J(\mathbf{r})$ and it does not imply a translational symmetry breaking spatial inhomogeneity of the spin exchange coupling $J$'s.Furthermore, the displacement vector $\Delta \mathbf{r}$ satisfies boundary conditions defined in parameter space such that $\Delta\mathbf{r}(\{J_{ij}\})=0$ when $\{J_{ij}\}$ gives the ordered state $A$ and $\Delta\mathbf{r}(\{J_{ij}\})=\mathbf{r}_{j}-\mathbf{r}_i$ (lattice site vectors) when $\{J_{ij}\}$ gives the ordered state $B$.

The gradient expansion Eq.(\ref{expansion}) will underlie the derivation of the action (free energy) difference that describes the intermediate phase in which an emergent Chern-Simons term appears naturally.The details of the derivation, which starts from the effective Hamiltonians Eqs.(\ref{lowenergyHamiltonian}-\ref{GSenergy}) accompanied by the gradient expansion Eq.(\ref{expansion}), are given in Appendix B.The main result is that the action (free energy) difference is given by
\[
S_{CSL}[A]=S[\phi^B]-S[\phi^A]=S_0[\phi^B]-S_0[\phi^A]
\]
\begin{widetext}
\[
\fl +i\frac{1}{2}S^2\int d\tau\sum_{ij}J_{ij}\cos\Delta\phi^0_{ij}\sin\theta^0_i\sin\theta^0_j
[(\Delta\mathbf{r}_{ij})^2(2\Delta \tau (\partial_x\Delta y)A^A_y\partial_{\tau} A^A_x+2\Delta \tau (\partial_y\Delta x) A^A_x\partial_{\tau} A^A_y)
\]
\[
\fl -(2\Delta \tau [ (\Delta x_{ij}(\partial_x\Delta y_r)-\Delta y_{ij}(\partial_x\Delta x_r))\Delta x_{ij} A^A_x \partial_{\tau} A^A_y+(\Delta y_{ij}(\partial_y\Delta x_r)-\Delta x_{ij}(\partial_y\Delta y_r))\Delta y_{ij} A^A_y \partial_{\tau} A^A_x ])
\]
\begin{equation}\label{CSlattice}
+(\Delta\tau)^2 ((A^B_0)^2-(A^A_0)^2)+\cdots]
\end{equation} 
\end{widetext}
where $S_0[\phi^{\alpha}]=\int d\tau \left(\phi^{\alpha}\partial_{\tau}\phi^{\alpha}+H_0[\phi^{\alpha}]\right)$ in Euclidean space-time with its imaginary time $\tau=i t$ and $H_0[\phi^{\alpha}]$ given by Eq.(\ref{GSenergy}).The $\cdots$ represents the terms containing the gauge field $A_{\mu}$ coming from those in Eq.(\ref{gradientexpansion}) but of no particular interest for this discussion.

Apart from the Berry phase and static configuration energy in the first term, the rest of Eq.(\ref{CSlattice}) are of 2+1-D Chern-Simons form, which in the gauge where $A_0=0$ and in the continuum limit give rise to a topological action of the form
\begin{equation}\label{CSterm}
S_{CS}[A]=i \int d^3 x \kappa(A_x \partial_{\tau} A_y - A_y\partial_{\tau} A_x)=i \int d^3 x\kappa\epsilon^{\mu\nu\lambda}A_{\mu}\partial_{\nu}A_{\lambda}
\end{equation}
in Euclidean space-time, where $A_{\mu}=\partial_{\mu}\phi$ with the geometric prefactors in Eqs.(\ref{gaugefield}) and (\ref{CSlattice}) absorbed into the coupling constant $\kappa$ for simplicity.Other than a constraint on the $\partial_{x,y}\Delta x_r, \partial_{x,y}\Delta y_r$ which reflects the required geometry and interaction-induced frustration effects, the above result is obtained only if the following `matching condition',
\begin{equation}\label{criterion2}
|\cos\Delta\phi^0_{ij}\sin\theta^0_i\sin\theta^0_j|_A=|\cos\Delta\phi^0_{ij}\sin\theta^0_i\sin\theta^0_j|_B
\end{equation}
is satisfied between two ordered ground states $A$ and $B$, as can be easily deduced from Eq.(\ref{CSlattice}).This means only appropriate two ordered ground states can give rise to Chern-Simons effective theory describing a gapped chiral spin liquid at the phase boundary between them.In considering the three ordered states; cuboc1, $\mathbf{q}=0$, and $\sqrt{3}\times\sqrt{3}$ around the $J_2=J_3=0$ point found in \cite{Messio}, we found the following results for the comparison of $h_{ij}=|\cos\Delta\phi^0_{ij}\sin\theta^0_i\sin\theta^0_j|$.
\begin{widetext}
\begin{center}
    \begin{tabular}{ | l | l | l | p{5cm} |}
    \hline
    ordered state & $h_{ij}$ ($1^\mathrm{st}$ neighbor) & $h_{ij}$ ($2^{\mathrm{nd}}$ neighbor) & $h_{ij}$ ($3^{\mathrm{rd}}$ neighbor) \\ \hline
    cuboc1 & $\frac{1}{2}$  & $\frac{1}{2}$ & 1 \\ \hline
    $\mathbf{q}=0$ & $\frac{1}{2}$  & $\frac{1}{2}$ & 1 \\ \hline
    $\sqrt{3}\times\sqrt{3}$ & $\frac{1}{2}$  & 1 & $\frac{1}{2}$ \\
    \hline
%\caption{Table 1.The comparison of $h_{ij}$ between the ordered ground states of kagom\'{e} antiferromagnet.}
    \end{tabular}
\end{center}
Table 1.The comparison of $h_{ij}$ between the ordered ground states of kagom\'{e} antiferromagnet with further neighbor spin exchange interactions.
\end{widetext}

From the Table 1., we clearly see that the function $h_{ij}$ is precisely equal between that for the cuboc1 and $\mathbf{q}=0$ states, while they are different between the cuboc1 and $\sqrt{3}\times\sqrt{3}$ states or between the $\mathbf{q}=0$ and $\sqrt{3}\times\sqrt{3}$ states.As a result, one obtains the Chern-Simons term Eq.(\ref{CSterm}) exactly at the phase transition between the cuboc1 and $\mathbf{q}=0$ states.We would like to argue that this is the explanation for why numerical works found the chiral spin liquid to arise at the phase transition between $\mathbf{q}=0$ and cuboc1 states but not between $\sqrt{3}\times\sqrt{3}$ and cuboc1 transition or $\sqrt{3}\times\sqrt{3}$ and $\mathbf{q}=0$ transition \cite{Messio}\cite{DNShengSciRepFQHEkagome}\cite{Globalphasediagram}\cite{Lauchli}.

This result is also nicely in agreement with the numerical result of the $J_1-J_2$ model of Heisenberg kagom\'{e} where the ground states are ferromagnet for $J_2/|J_1|<1/3$ and cuboc for $J_2/|J_1|>1/3$ \cite{Domenge} for which our matching condition can never be satisfied for such ground states since the ferromagnet order can be chosen to point in the $x$ direction for example, giving $h_{ij}=1$ for both first and second neighbor whereas the cuboc state gives rise to $h_{ij}\neq 1$ for both first and second neighbors.Thus, the Chern-Simons effective theory cannot appear at the ferromagnet-cuboc transition in the $J_1-J_2$ kagom\'{e} Heisenberg model, explaining the absence of chiral spin liquid in such model \cite{Domenge}.This, in combination with Eq.(\ref{CSBterm}) from our analysis based on Baskaran's chiral order parameter theory, explains the conclusion of numerous numerical results on why kagom\'{e} antiferromagnet with third neighbor spin exchange interaction apparently gives rise to a chiral spin liquid ground state while with only nearest-neighbor \cite{White}\cite{Schollwock} and second neighbor spin exchanges \cite{Domenge}, no such chiral spin liquid was found.Based on Eqs.(\ref{gaugefieldHamiltoniancuboc1result}) and (\ref{weightcuboc}) in conjunction with Eq.(\ref{CSBterm}), this comes from the dominance of the third neighbor term to the chiral symmetry breaking effect that gives rise to global Ising $\mathbb{Z}_2$ symmetry needed to have chiral spin liquid.It is interesting to note that our result is also in agreement with a recent numerical work where a similar chiral spin liquid state was found to occur in the $J_1-J_2-J_3$ kagom\'{e} Heisenberg model, but which is gapless because of the ferromagnetic first and second neighbor spin exchange interactions \cite{LPMC}. 
  
We thus have derived explicitly the emergent Chern-Simons theory that describes the chiral spin liquid at the phase transition between two ordered ground states in kagom\'{e} antiferromagnet, as concluded numerically from recent works \cite{Messio}\cite{DNShengSciRepFQHEkagome}\cite{Lauchli}.Such Chern-Simons field theory is of high significance in the study of quantum field theory and gravity and represents a mass term for the gauge field \cite{DeserJackiwTempleton}\cite{Schoenfeld}.Our derivation above is also in agreement with the observation that Chern-Simons term can be generated not only from radiative correction \cite{Dunne} as is established in the study of topologically massive gauge theory \cite{DeserJackiwTempleton}\cite{Schoenfeld} but also from spontaneous symmetry breaking \cite{KhareReview}.In this case, the spontaneous symmetry breaking in the neighboring cuboc1 and $\mathbf{q}=0$ ordered states gives rise to the Chern-Simon effective theory at the phase transition between them.

\section{Conclusion}
We have shown that further neighbor spin exchange interactions in kagom\'{e} antiferromagnet give rise to significant quantum fluctuations as strong as those from first neighbor.In particular, third neighbor spin exchange interaction in kagom\'{e} antiferromagnet is special, not only that it produces strong quantum fluctuations, but also chiral symmetry-breaking effect, which we deduced using a chiral order parameter theory.More precisely, while finite temperature chiral ordered state exists in $J_1-J_2$ kagom\'{e} Heisenberg model, one needs to go at least to third neighbor, that is $J_1-J_2-J_3$ model to get chiral spin liquid ground state.

We have also demonstrated that Chern-Simons effective theory appears naturally at the transition between two ordered ground states satisfying appropriate matching condition, which is satisfied and explains the emergence of chiral spin liquid in $J_1-J_2-J_3$ kagom\'{e} antiferromagnet while it is not satisfied and chiral spin liquid is thus absent in $J_1$-only and $J_1-J_2$-only models.It is an interesting open problem to find the description of energy spectrum (eigenstates structure) of the chiral spin liquid, such as the gapped energy spectrum of its spin excitations within this field theoretical picture.It is also to be noted that even though our calculations are applied to a specific model and the associated numerical result, our formalism is completely general and should be implementable to other 2d frustrated spin systems.

It is instructive to compare or connect our theory with the available experiments on real magnetic materials with kagom\'{e} lattice structure.A possible candidate from such materials where a chiral spin state may occur is the $S=5/2$ potassium jarosite $\mathrm{KFe_3(OH)_6(SO_4)_2}$.An early experimental work using nuclear magnetic resonance (NMR) on powder samples of this compound \cite{jarosite} already identified and confirmed chiral ordered spin state with definite spin chirality (either $+1$ or $-1$) at low temperatures associated with the $\mathbf{q}=0$ ordered state while such definite chirality spin state does not occur for the $\sqrt{3}\times\sqrt{3}$ state as the latter contains both the $\pm 1$ chiralities, in perfect consistency with the conclusion of our analysis in Section 3.In this work \cite{jarosite}, the chiral ordered state at low but finite temperatures is associated with Dzyaloshinskii-Moriya interaction and spin exchange anisotropy.Spin chirality in the same compound but in large single-crystal samples described by Heisenberg model has been measured in \cite{spinchirality}, where a combination of thermodynamic and neutron scattering analyses was employed.Just like in \cite{jarosite}, a Heisenberg model with only nearest-neighbor (antiferromagnetic) spin exchange interaction was assumed to analyze the experimental results.Some indications of chiral spin state at low temperatures were observed but they were also ascribed to Dzyaloshinskii-Moriya interaction.While the origin of the spin chirality in our work is further-neighbor spin exchange interactions rather than such an asymmetric spin exchange interaction as Dzyaloshinskii-Moriya coupling, the observed presence of chiral spin order in those works \cite{jarosite}\cite{spinchirality} is already in tune with our theoretical results  described in Section 3 that argues for the occurrence of chiral spin liquid in kagom\'{e} Heisenberg antiferromagnet.In fact, we would like to suggest that should further-neighbor spin exchange interactions in the jarosite family of compounds $R\mathrm{(Fe/Cr)_3(OH)_6(SO_4)_2}$ (where $R=\mathrm{K,Na,NH_4,etc.}$) be strong enough while anisotropy or Dzyaloshinskii-Moriya interaction be negligible or much weaker, our theory for the chiral spin ordered state at low temperatures could provide an alternative explanation for the occurrence of chiral spin states at low temperatures and further implies a striking prediction that at the absolute zero temperature, a chiral spin liquid state may actually occur (between such ordered states) if the microscopic coupling parameters are just right to permit this latter exotic state (that is, if the spin exchange couplings place the compound to be around the boundary between the $\mathbf{q}=0$ and cuboc1 ordered ground states, the latter of which occurs with further spin exchange interaction).This should motivate further experimental works on this topic.     

\bigskip
 
\textit{Acknowledgements.\textemdash}I.M. thanks Prof. T. Roscilde, Prof. A. Fedorenko, Prof. K. Gawedzki, and Prof. H. Samtleben for the very helpful discussions and Prof. M. Vojta for the constructive criticisms that have led to the significant improvement of the manuscript.

\bigskip

%\bibliography{sample.bib}






% \end{thebibliography}

\appendix

\section{Derivation of the Free Energy in the Chiral Order Parameter Theory}
In this appendix, we give the details of the derivation of the free energy terms in the chiral order parameter theory in Eqs.(\ref{CSBterm}-\ref{chiralitycorrelationfunction}) using the method of computing a correlation function standard in quantum field theory or many-body physics.

The main quantities to compute according to Eq.(\ref{chiralitycorrelationfunction}) are the correlation functions $\langle \mathcal{T}\left[\chi^{\bigtriangleup}_{ijk}(\tau_1)\chi^{\bigtriangledown}_{jlm}(\tau_2)\right]\rangle$ for the second neighbor and $\langle \mathcal{T}\left[\chi^{\frac{\bigtriangleup}{\bigtriangledown}}_{ijk}(\tau_1)\chi^{\frac{\bigtriangleup}{\bigtriangledown}}_{mln}(\tau_2)\right]\rangle$ for the third neighbor spin exchange interactions respectively.For this purpose, we will heavily use Green's function approach.The Green's function can be deduced from Eq.(\ref{Hamiltonianfluctuation}), which gives
\begin{widetext}
\[
\fl G^{-1}_{\alpha,\beta}(\mathbf{r},\mathbf{r}',\tau) =\frac{1}{2}S^2J_{ij}\delta(|\Delta\mathbf{r}|-|\Delta\mathbf{r}_{ij}|)\times
\]
\begin{equation}\label{GF}
 \left( \begin{array}{cc}
\cos\Delta\phi^0_{\mathbf{r}\mathbf{r}'}\sin\theta^0_{\mathbf{r}}\sin\theta^0_{\mathbf{r}'} & \sin\theta^0_{\mathbf{r}}\cos\theta^0_{\mathbf{r}'}\sin\Delta\phi^0_{\mathbf{r}\mathbf{r}'} \\
\sin\theta^0_{\mathbf{r}}\cos\theta^0_{\mathbf{r}'}\sin\Delta\phi^0_{\mathbf{r}\mathbf{r}'}&\cos\Delta\phi^0_{\mathbf{r}\mathbf{r}'}\sin\theta^0_{\mathbf{r}}\sin\theta^0_{\mathbf{r}'} +\cos\theta^0_{\mathbf{r}}\cos\theta^0_{\mathbf{r}'}
\end{array} \right)(\nabla^2+\partial^2_{\tau})
\end{equation}
\end{widetext}
where $\Delta\mathbf{r}=\mathbf{r}'-\mathbf{r}$.

In terms of the phase variables, the scalar spin chirality in Eq.(\ref{scalarspinchiralityfluctuation}) is given by
\begin{widetext}
\[
\fl \delta\chi_{ijk}(\tau)=
\]
\[
\fl     \delta \theta(\tau) S^3 (\cos
    \phi^0_j (\cos[\theta^0_i + \theta^0_k] (\Delta s_{ki}) \sin\theta^0_j 
         +   \cos\theta^0_j (-\cos\theta^0_k \sin\phi^0_i \sin\theta^0_i + 
                 \cos\theta^0_i \sin\phi^0_k \sin\theta^0_k)) 
         \]
         \[
\fl+   \cos\phi^0_k (\cos
        \theta^0_j (\cos\theta^0_k \sin\phi^0_i \sin\theta^0_i + 
         \cos\theta^0_i (\Delta s_{ij}) \sin\theta^0_k) 
           - 
      \sin\theta^0_j (\cos\theta^0_i \cos\theta^0_k \sin
           \phi^0_j + (\Delta s_{ij}) \sin\theta^0_i \sin\theta^0_k)) 
           \]
 \begin{equation}\label{scalarspinchiralityfluctuations}
\fl           + 
   \cos\phi^0_i (\cos
        \theta^0_j (\cos\theta^0_k (\Delta s_{jk}) \sin\theta^0_i
              - 
         \cos\theta^0_i \sin\phi^0_k \sin\theta^0_k) 
        + 
      \sin\theta^0_j (\cos\theta^0_i \cos\theta^0_k \sin
           \phi^0_j + (-\Delta s_{jk}) \sin\theta^0_i \sin\theta^0_k)))
\end{equation}
\end{widetext}
with $\Delta s_{ij}=\sin\phi^0_i - \sin\phi^0_j$ where we note that, as expected in working to order $\mathcal{O}(\delta S)$ for the $\delta\chi_{ijk}$, the fluctuations of the scalar spin chirality depend only on the polar angle fluctuations $\delta\theta$, since the in plane fluctuations proportional to $\delta\phi$ has zero projection in the normal-to-the-plane $z$ direction.This result holds even for out-of-plane spins that are present for example in cuboc1 state.

We can re-write the fluctuation function of the scalar spin chirality more concisely $\delta\chi_{ijk}(\tau)=\delta\theta(\tau)S^3 f_{ijk}(\phi^0,\theta^0)$ where the function $f_{ijk}(\phi^0,\theta^0)$, to be called the scalar spin chirality `form factor', is given by the expression on the right hand side of Eq.(\ref{scalarspinchiralityfluctuations}).It is to be noted that this form factor is static time-independent quantity given an ordered state.For coplanar ordered state, $\theta^0=\pi/2$ at all sites and the chirality form factor simplifies to
\begin{widetext}
\begin{equation}
\fl f_{ijk}(\phi^0,\theta^0=\frac{\pi}{2})=S^3 (\cos\phi^0_k (-\sin\phi^0_i + \sin\phi^0_j) + 
   \cos\phi^0_j (\sin\phi^0_i - \sin\phi^0_k) + 
   \cos\phi^0_i (-\sin\phi^0_j + \sin\phi^0_k))
\end{equation}
\end{widetext}
The scalar spin chirality correlation is then given by
\begin{widetext}
\begin{equation}\label{NNNchiralitycorrelfunction}
\fl \langle \mathcal{T}\left[\chi^{\bigtriangleup}_{ijk}(\tau_1)\chi^{\bigtriangledown}_{jlm}(\tau_2)\right]\rangle=S^6f^{\bigtriangleup}_{ijk}(\phi^0,\theta^0)f^{\bigtriangledown}_{jlm}(\phi^0,\theta^0)\frac{\int d\zeta \mathcal{T}\left[e^{-\int d\tau \left[\zeta^T(\partial_{\tau}+H[\phi,\theta])\zeta+J(\tau)\delta\theta(\tau)\right]}\delta\theta(\tau_1)\delta\theta(\tau_2)\right]}{\int d\zeta e^{-\int d\tau \left[\zeta^T(\partial_{\tau}+H[\phi,\theta])\zeta+J(\tau)\delta\theta(\tau)\right]}}|_{J=0}
\end{equation}
\begin{equation}\label{NNNNchiralitycorrelfunction}
\fl \langle \mathcal{T}\left[\chi^{\bigtriangleup}_{ijk}(\tau_1)\chi^{\bigtriangleup}_{mln}(\tau_2)\right]\rangle=S^6f^{\bigtriangleup}_{ijk}(\phi^0,\theta^0)f^{\bigtriangleup}_{mln}(\phi^0,\theta^0)\frac{\int d\zeta \mathcal{T}\left[e^{-\int d\tau \left[\zeta^T(\partial_{\tau}+H[\phi,\theta])\zeta+J(\tau)\delta\theta(\tau)\right]}\delta\theta(\tau_1)\delta\theta(\tau_2)\right]}{\int d\zeta e^{-\int d\tau \left[\zeta^T(\partial_{\tau}+H[\phi,\theta])\zeta+J(\tau)\delta\theta(\tau)\right]}}|_{J=0}
\end{equation} 
\end{widetext}
for the second and third neighbor exchange-induced correlations respectively.We then obtain
%\begin{widetext}
\[
\langle\mathcal{T}\left[\delta\theta(\tau_1)\delta\theta(\tau_2)\right]\rangle=
\frac{\int d\zeta \mathcal{T}\left[e^{-\int d\tau \left[\zeta^T\left(\partial_{\tau}+H[\phi,\theta]\right)\zeta+J(\tau)\delta\theta(\tau)\right]}\delta\theta(\tau_1)\delta\theta(\tau_2)\right]}{\int d\zeta e^{-\int d\tau \left[\zeta^T(\partial_{\tau}+H[\phi,\theta])\zeta+J(\tau)\delta\theta(\tau)\right]}}|_{J=0}
\]
\begin{equation}\label{chiralspincorrelation}
=\frac{1}{i^2}\frac{\delta^2 \mathcal{Z}[J]}{\delta J(\tau_1)\delta J(\tau_2)}|_{J=0}
\end{equation}
%\end{widetext}
where we have used standard result for n-point correlation function \cite{ZinnQFT}\cite{RyderQFT} and with
%\begin{widetext}
\[
 \mathcal{Z}[J]=
 \frac{\int d\zeta e^{-\int d\tau \left[\zeta^T(\partial_{\tau}+H[\phi,\theta])\zeta+J(\tau)\delta\theta(\tau)\right]}}{\int d\zeta e^{-\int d\tau \left[\zeta^T(\partial_{\tau}+H[\phi,\theta])\zeta\right]}}
=\frac{\int d\zeta e^{-\int d\tau \left[\zeta^TM[\theta(\tau)]\zeta+J(\tau)\delta\theta(\tau)\right]}}{\int d\zeta e^{-\int d\tau \left[\zeta^TM[\theta(\tau)]\zeta\right]}}
\]
%\end{widetext}
\begin{equation}\label{Mmatrix}
M[\theta(\tau)]=\mathbb{I}_{2\times 2}\partial_{\tau}+G^{-1}[\theta(\tau)]
\end{equation}
where $G^{-1}[\theta(\tau)]$ is given in Eq.(\ref{GF}).Using Eqs.(\ref{NNNchiralitycorrelfunction}-\ref{Mmatrix}), we have
%\begin{widetext}
\begin{equation}\label{chiralitycorrfunctiontemporalapp}
\fl \langle \mathcal{T}\left[\chi_{ijk}(\tau_1)\chi_{mln}(\tau_2)\right]\rangle=S^6f_{ijk}(\phi^0,\theta^0)f_{mln}(\phi^0,\theta^0)\frac{1}{2}\int_{\mathbf{x}_1}\int_{\mathbf{x}_2}\frac{1}{\partial_{\tau}+\tilde{g}^{-1}_{\mathbf{x}_1\mathbf{x}_2}(\partial^2_{\tau}+\nabla^2)}
\end{equation}
%\end{widetext}
with $\tau=\tau_2-\tau_1$ and in which
\begin{widetext} 
\begin{equation}\label{gfunctionapp}
\fl \tilde{g}^{-1}_{\mathbf{x}_1\mathbf{x}_2}=\frac{1}{2}S^2J_{ij}\delta(|\Delta\mathbf{x}_{12}|-|\Delta\mathbf{r}_{ij}|)\left[\cos\Delta\phi^0_{\mathbf{x}_1\mathbf{x}_2}\sin\theta^0_{\mathbf{x}_1}\sin\theta^0_{\mathbf{x}_2} +\cos\theta^0_{\mathbf{x}_1}\cos\theta^0_{\mathbf{x}_2}\right]
\end{equation}
\end{widetext}
where $\Delta\mathbf{x}_{12}=\mathbf{x}_2-\mathbf{x}_1$ and for this calculation, according to Eq.(\ref{chiralitycorrelationfunction}), we have $|\Delta\mathbf{r}_{nnn}|=\sqrt{3},|\Delta\mathbf{r}_{nnnn}|=2$ for the second and third neighbor spin exchanges, respectively.We evaluate the above result Eqs.(\ref{chiralitycorrfunctiontemporalapp}-\ref{gfunctionapp}) for the coplanar $\mathbf{q}=0$ and $\sqrt{3}\times\sqrt{3}$ states where $\Delta\phi^0_{ij\in nnn}=2\pi/3,\Delta\phi^0_{ij\in nnnn}=0$ and $\Delta\phi^0_{ij\in nnnn}=0,\Delta\phi^0_{ij\in nnnn}=2\pi/3$ respectively and also for the cuboc1 state, we obtain
\begin{widetext}
\begin{equation}\label{gfactorq=0}
\fl \tilde{g}^{-1}_{\mathbf{x}_1\mathbf{x}_2}|_{\mathbf{q}=0}=-\frac{1}{2}S^2\left(\frac{1}{2}|J_2|\delta(|\mathbf{x}_2-\mathbf{x}_1|-|\Delta\mathbf{r}_{nnn}|)-|J_3|\delta(|\mathbf{x}_2-\mathbf{x}_1|-|\Delta\mathbf{r}_{nnnn}|)\right)
\end{equation}
\begin{equation}\label{gfactor3x3}
\fl \tilde{g}^{-1}_{\mathbf{x}_1\mathbf{x}_2}|_{\sqrt{3}\times\sqrt{3}}=-\frac{1}{2}S^2\left(|J_2|\delta(|\mathbf{x}_2-\mathbf{x}_1|-|\Delta\mathbf{r}_{nnn}|)+\frac{1}{2}|J_3|\delta(|\mathbf{x}_2-\mathbf{x}_1|-|\Delta\mathbf{r}_{nnnn}|)\right)
\end{equation}
\begin{equation}\label{gfactorcuboc1}
\fl \tilde{g}^{-1}_{\mathbf{x}_1\mathbf{x}_2}|_{\mathrm{cuboc1}}=\frac{1}{2}S^2\left(\zeta|J_2|\delta(|\mathbf{x}_2-\mathbf{x}_1|-|\Delta\mathbf{r}_{nnn}|)-|J_3|\delta(|\mathbf{x}_2-\mathbf{x}_1|-|\Delta\mathbf{r}_{nnnn}|)\right)
\end{equation}
\end{widetext}
where we have made use of the fact that numerical calculations show that the $\mathbf{q}=0$-cuboc1 phase transition occurs for $J_2>0,J_3>0$ quadrant of the phase diagram while the $\sqrt{3}\times\sqrt{3}$-cuboc1 phase transition occurs in the $J_2<0,J_3>0$ quadrant of the phase diagram whereas cuboc1 state occurs for $J_3>0$ and $J_2>0(J_2<0)$ at the side bordering with the $\mathbf{q}=0(\sqrt{3}\times\sqrt{3})$ state, giving $\zeta=+1(-1)$ respectively.The three transitions meet at the $J_2=J_3=0$ tricritical point \cite{Messio}.This is the result of classical spin calculation.In the quantum calculation, the phase diagram shifts the triciritical point slightly away from the $J_2=J_3=0$ point \cite{Messio}.From the above result Eqns.(\ref{gfactorq=0}) and (\ref{gfactor3x3}), we note that the $\tilde{g}^{-1}_{\mathbf{x}_1\mathbf{x}_2}|_{\mathbf{q}=0}$ has both negative-valued and positive-valued components, coming from  $J_2$ and $J_3$ contributions respectively, whereas $\tilde{g}^{-1}_{\mathbf{x}_1\mathbf{x}_2}|_{\sqrt{3}\times\sqrt{3}}$ is always negative in sign, regardless of the relative magnitude of $|J_2|$ and $|J_3|$.On the other hand, very interestingly, for the noncoplanar cuboc1 state, the second and third neighbor contributions are of precisely the same weight but opposite in sign but what is important is it does have positive component, which comes from second neighbor contribution at the border between the cuboc1 and $\mathbf{q}=0$ state.This will have crucial consequence as we will show later.In the main text, we analyzed the physical implication of the above result for the three ordered states of the kagom\'{e} Heisenberg model with up to third neighbor isotropic spin exchange interactions that we considered in this work. 

The (imaginary) time-dependence of the chirality-chirality correlation function is determined by the $\partial_{\tau}$ and $\partial^2_{\tau}$ terms in the Eq.(\ref{chiralitycorrfunctiontemporal}).We note from the Eq.(\ref{chiralitycorrfunctiontemporal}) that the net dependence on this (imaginary bosonic Matsubara) frequency $k_0=2\pi n T$ takes the form $\sim 1/(ik_0+\tilde{g}^{-1}_{x_1x_2}(-k^2_0+\nabla^2))$ which at $T=0$ gives rise to temporal dependence
\begin{widetext}
\[
\langle \mathcal{T}\left[\chi_{ijk}(\tau_1)\chi_{mln}(\tau_2)\right]\rangle\equiv\langle \mathcal{T}\left[\chi_{ijk}(0)\chi_{mln}(\tau)\right]\rangle
\]
\[
= S^6f_{ijk}(\phi^0,\theta^0)f_{mln}(\phi^0,\theta^0)\frac{1}{2}\int_{\mathbf{x}_1}\int_{\mathbf{x}_2}\frac{1}{\partial_{\tau}+\tilde{g}^{-1}_{\mathbf{x}_1\mathbf{x}_2}(\partial^2_{\tau}+\nabla^2)}
\]
\[
= S^6f_{ijk}(\phi^0,\theta^0)f_{mln}(\phi^0,\theta^0)\frac{1}{2}\int_{\mathbf{x}_1}\int_{\mathbf{x}_2}\int \frac{dk_0}{2\pi}\frac{e^{ik_0(\tau_2-\tau_1)}}{ik_0+\tilde{g}^{-1}_{\mathbf{x}_1\mathbf{x}_2}(-k^2_0+\nabla^2)}
\]
\begin{equation}\label{chiralitycorrfunctiontemporalfinal}
= S^6f_{ijk}(\phi^0,\theta^0)f_{mln}(\phi^0,\theta^0)\frac{1}{2}\int_{\mathbf{x}_1}\int_{\mathbf{x}_2}F_{\mathbf{x}_1\mathbf{x}_2}(\tau_2-\tau_1)
\end{equation}
\end{widetext}
where 
%\begin{widetext}
\[
F_{\mathbf{x}_1\mathbf{x}_2}(\tau)=-\frac{\omega_0e^{\omega_0\tau} \sqrt{2\pi}}{\omega_{\mathbf{k}}} (e^{ -\omega_{\mathbf{k}} \tau}\Theta[\tau \mathcal{S}[\omega_{\mathbf{k}} - \omega_0]] \mathcal{S}[\omega_{\mathbf{k}} - \omega_0] 
\]
\begin{equation}\label{temporalcorrelation}
+e^{ \omega_{\mathbf{k}} \tau}
        \Theta[-\tau \mathcal{S}[\omega_{\mathbf{k}} + \omega_0]] \mathcal{S}[\omega_{\mathbf{k}} + \omega_0]))
\end{equation}
%\end{widetext}
where $\mathcal{S}(\cdots)$ refers to the sign of $(\cdots)$ and 
\begin{equation}\label{coefficientsinF}
\omega_0=\frac{1}{2\tilde{g}^{-1}_{\mathbf{x}_1\mathbf{x}_2}},\omega^2_{\mathbf{k}}=\frac{1}{4(\tilde{g}^{-1}_{\mathbf{x}_1\mathbf{x}_2})^2}-\nabla^2 =\mathrm{I.F.T}\left[\frac{1}{4}\mathcal{G}_{\mathbf{k}}+\mathbf{k}^2\right]
\end{equation}
with $\Theta[z]$ the Heaviside theta function; $\Theta[z]=1$ for $z\geq 0$ and zero otherwise and we have symbolically labeled the (spatial) Fourier transform by $\mathrm{F.T.}$(e.g. $\mathcal{G}_{\mathbf{k}}=\mathrm{F.T.}\left[1/(\tilde{g}^{-1}_{\mathbf{x}_1\mathbf{x}_2})^2\right]$) and its inverse by $\mathrm{I.F.T}$.

\section{Derivation of the Emergent Chern-Simons Term}
The derivation of the action (free energy) difference that gives rise to an emergent Chern-Simons term starts from standard definition $S[\phi^{\alpha}]=\int d\tau \left(\phi^{\alpha}\partial_{\tau}\phi^{\alpha}+H[\phi^{\alpha}]\right)$ in Euclidean space-time with its imaginary time $\tau=i t$.In this case, the Hamiltonian is given by $H[\phi^{\alpha}]=H_0[\phi^{\alpha}]+\delta H[\phi^{\alpha}]$ where $H_0[\phi^{\alpha}]$ from Eq.(\ref{GSenergy}) corresponds to the ground state energy while $\delta H[\phi^{\alpha}]$ from Eq.(\ref{lowenergyHamiltonian}) corresponds to the quantum fluctuations energy, aided with the gradient expansion relation Eq.(\ref{expansion}).We will drop the lattice index of the position vector in the continuum limit.Taking the gradient of both sides of the equation (with respect to $\mathbf{r}$), we obtain
%\begin{widetext}
\[
\nabla\phi^B(\mathbf{r},t)=\nabla\phi^A(\mathbf{r},t)+\nabla\left[\Delta\mathbf{r}\cdot\nabla\phi^A(\mathbf{r},t)\right]
\]
\begin{equation}
+\nabla[\Delta t\partial_t\phi^A(\mathbf{r},t)]+\mathcal{O}(|\Delta\mathbf{r}|^2,(\Delta t)^2,|\Delta\mathbf{r}|\Delta t)
\end{equation}
%\end{widetext}
Using vector identity
\begin{equation}
\nabla(\mathbf{a}\cdot\mathbf{b})=(\mathbf{a}\cdot\nabla)\mathbf{b}+(\mathbf{b}\cdot\nabla)\mathbf{a}+\mathbf{a}\times(\nabla\times\mathbf{b})+\mathbf{b}\times(\nabla\times\mathbf{a})
\end{equation}
we obtain
\begin{widetext}
\begin{equation}
\fl\nabla\left[\Delta\mathbf{r}\cdot\nabla\phi^A(\mathbf{r},t)\right]=(\Delta\mathbf{r}\cdot\nabla)\nabla\phi^A+(\nabla\phi^A\cdot\nabla)\Delta\mathbf{r}+\Delta\mathbf{r}\times(\nabla\times\nabla\phi^A)+\nabla\phi^A\times(\nabla\times\Delta\mathbf{r})
\end{equation}
\end{widetext}
If we work up to first order in $\nabla\phi^A$, then we can keep the second and fourth terms in the above equation and omit the other two higher order terms.We will make use of the useful identity 
\begin{equation}\label{identity}
(\nabla\phi\cdot\Delta\mathbf{r}_{ij})^2=(\Delta\mathbf{r}_{ij})^2(\nabla\phi)^2-(\nabla\phi\times\Delta\mathbf{r}_{ij})^2
\end{equation}
Now consider the gradient of the phase field squared,
\begin{widetext}
\[
\fl (\nabla\phi^B(\mathbf{r},t))^2=(\nabla\phi^A(\mathbf{r},t))^2+\cdots+2\left[(\nabla\phi^A\cdot\nabla)\Delta\mathbf{r}+\nabla\phi^A\times(\nabla\times\Delta\mathbf{r})\right]\cdot\Delta t\partial_t\nabla\phi^A(\mathbf{r},t)+\cdots
\]
\end{widetext}
and using $A^{a=A,B}_i=\partial\phi^{a=A,B}_i$ where $i=x,y$ and the geometric prefactor in Eq.(\ref{gaugefield}) to be absorbed into a coupling constant defined later on, we obtain
%\begin{widetext}
\[
(\mathbf{A}^B)^2=(\mathbf{A}^A)^2+\cdots+v_xA^A_x\partial_tA^A_x+v_yA^A_y\partial_tA^A_x
\]
\begin{equation}\label{gradientexpansion}
+w_xA^A_x\partial_tA^A_y+w_yA^A_y\partial_tA^A_y+\cdots
\end{equation}
%\end{widetext}
where $v_{x(y)},w_{x(y)}$ are constant coefficients and are given by
\[
v_x=2\Delta t (\partial_x\Delta x_r),v_y=2\Delta t (\partial_x\Delta y_r),
\]
\begin{equation}\label{coefficients}
w_x=2\Delta t (\partial_y\Delta x_r),w_y=2\Delta t (\partial_y\Delta y_r),
\end{equation}
taken in the continuum limit where the above products take fixed values.We note the presence of the terms
%\begin{widetext}
\[
v_yA^A_y\partial_tA^A_x+w_xA^A_x\partial_tA^A_y=
\]
\begin{equation}\label{CStermfromdot}
2\Delta t (\partial_x\Delta y_r)A^A_y\partial_t A^A_x+2\Delta t (\partial_y\Delta x_r) A^A_x\partial_t A^A_y
\end{equation}
%\end{widetext}

On the other hand, the $(\nabla\phi\times\Delta\mathbf{r}_{ij})^2$ term in Eq.(\ref{identity}) also gives a Chern-Simons type term as
%\begin{widetext}
\[
2\Delta t [ (\Delta x_{ij}(\partial_x\Delta y_r)-\Delta y_{ij}(\partial_x\Delta x_r))\Delta x_{ij} A^A_x \partial_t A^A_y
\]
\begin{equation}\label{CStermfromcross}
+(\Delta y_{ij}(\partial_y\Delta x_r)-\Delta x_{ij}(\partial_y\Delta y_r))\Delta y_{ij} A^A_y \partial_t A^A_x ]
\end{equation}
%\end{widetext}
where $\Delta\mathbf{r}_{ij}=(\Delta x_{ij},\Delta y_{ij})$.We substitute the above results Eqs.(\ref{CStermfromdot}-\ref{CStermfromcross}) into the standard expression for the action in Euclidean space-time $S[\phi^{\alpha}]=\int d\tau \left(\phi^{\alpha}\partial_{\tau}\phi^{\alpha}+H[\phi^{\alpha}]\right)$ with $H[\phi^{\alpha}]=H_0[\phi^{\alpha}]+\delta H[\phi^{\alpha}]$ using the Hamiltonians Eqs.(\ref{lowenergyHamiltonian}) and (\ref{GSenergy}) applied to the ordered state A with order parameter field $\phi^A$ and ordered state B with order parameter field $\phi^B$ and compute their difference.Performing this task, we obtain Eq.(\ref{CSlattice}) in the main text.

\end{document}